\documentclass[blind,review,3p]{elsarticle}

\usepackage{lineno,hyperref}
\modulolinenumbers[5]

\usepackage{amssymb}
\usepackage{mathtools}
\usepackage{setspace}
\usepackage{longtable}
\usepackage{threeparttable}
\usepackage{multicol}
\usepackage{multirow}
\usepackage{pdflscape}

\usepackage{tikz}
	\usetikzlibrary{arrows,chains,matrix,positioning,scopes}
	\tikzset{>=triangle 45} %latex or latex' or stealth or stealth'
\usepackage{pgfplots}
\usepackage{graphics}
	\graphicspath{{figures/}}
\usepackage{amsmath}
\usepackage{amssymb}
\usepackage{float}
\usepackage{multirow}
\usepackage{fourier} 
\usepackage{array}
\usepackage{makecell}

\usepackage{comment}

\newcommand{\indep}{\perp \!\!\! \perp}
\newcommand{\hquad}{\hspace{0.2em}}

	\usepackage{amsmath}
\usepackage{amssymb}
\usepackage{graphics}
	\graphicspath{{figures/}}
	
	\usepackage{lscape}
	\usepackage{pgfgantt}
	\usepackage{array,multirow}
	\usepackage{float}
	\usepackage{amsmath}
\usepackage{amssymb}
	
	\usepackage{multirow}
	\usepackage{numprint}

\newcommand{\Sigmabold}{\boldsymbol{\Sigma}}
\newcommand{\Omegabold}{\boldsymbol{\Omega}}

\usepackage{epstopdf}% To incorporate .eps illustrations using PDFLaTeX, etc.
\usepackage{subcaption}

\usepackage{lineno,hyperref}
\modulolinenumbers[5]

\usepackage{url}

\usepackage{graphics}
	\graphicspath{{figures/}}
	
	\usepackage{subcaption}
	\usepackage{lscape}
	\usepackage{pgfgantt}
	\usepackage{array,multirow}
	\usepackage{float}
\usepackage{amssymb}
	
	\usepackage{multirow}
	\usepackage{numprint}
\usepackage{comment}

\journal{arXiv}

\biboptions{authoryear}

\begin{document}

\begin{frontmatter}

\author[1]{Karoline Bax \corref{cor1}} %\fnref{label1}}
\author[1]{Emanuele Taufer}%\fnref{label2}}
\author[1]{Sandra Paterlini}%\fnref{label1}}

%\fntext[label1]{Department of Economics and Management, University of Trento, Trento, Italy}
%\fntext[label2]{Department of Mathematics, Technical University of Munich, Munich, Germany}

\address[1]{Department of Economics and Management, University of Trento, Trento, Italy}
%\address[2]{Department of Mathematics, Technical University of Munich, Munich, Germany}

\title{A generalized precision matrix for t-Student distributions in\\ portfolio optimization}

%% use optional labels to link authors explicitly to addresses:

\begin{abstract}
The Markowitz model is still the cornerstone of modern portfolio theory. 
In particular, when focusing on the minimum-variance portfolio, the covariance matrix or better its inverse, the so-called precision matrix, is the only input required. So far, most scholars worked on improving the estimation of the input, however little attention has been given to the limitations of the inverse covariance matrix when capturing the dependence structure in a non-Gaussian setting.  While the precision matrix allows to correctly understand the conditional dependence structure of random vectors in a Gaussian setting, the inverse of the covariance matrix might not necessarily result in a reliable source of information when Gaussianity fails. In this paper, exploiting the local dependence function,  different definitions of the generalized precision matrix (GPM), which holds for a general class of distributions,  are provided. In particular, we focus on the multivariate t-Student distribution and point out that the interaction in random vectors does not depend only on the inverse of the covariance matrix, but also on additional elements. We test the performance of the proposed GPM using a minimum-variance portfolio set-up by considering S\&P 100 and Fama and French industry data. We show that portfolios relying on the GPM often generate statistically significant lower out-of-sample variances than state-of-art methods.
\end{abstract}


\end{frontmatter}

\section{Introduction}\sloppy

Even after 70 years, the cornerstone of many sophisticated portfolio approaches is still centered around the risk-return optimization framework developed by \cite{markowitz1952portfolio}.  Especially, the minimum-variance framework is very popular due to its simplicity as the only input necessary for the analytical solution is the precision matrix, which is defined as the inverse of the covariance matrix in the Gaussian setting \citep{stevens1998inverse}.  More specifically,  when considering a  $d$-dimensional multivariate Gaussian vector  $\underline{X}=[X_1,X_2, \cdots, X_d]^{\top}$ with mean $\underline{\mu}$ and covariance $\Sigmabold$;   the precision matrix is defined as $\boldsymbol{\Omega}= \Sigmabold^{-1}$ and  $\omega_{pq}$ is its  $(p, q)$-th element.  

While most scholars focus on improving the estimation of the precision matrix as it received much criticism (see \cite{bloomfield1977portfolio, demiguel2009optimal,kritzman2010defense} for poor out-of-sample performance; \cite{michaud1989markowitz,ledoit2004honey,black1990asset, black1992global} for unreliable and extreme weights and \cite{ledoit2004well,meucci2009risk,won2013condition} for the ill-conditioning of the covariance matrix), not much attention has been put on discussing the flaws and limitations of the inverse  covariance matrix in correctly capturing the dependence structure when the Gaussian assumption fails. 

While in a Gaussian setting, the precision matrix can give information on the conditional independence, in fact  $\omega_{pq}=0$ if and only if  $X_p \indep X_q$ conditional on $\underline{X}_{\setminus (p,q)}$, where $\underline{X}_{\setminus (p,q)}$ indicates the vector $\underline{X}$ with the $(p,q)$-th elements removed \citep{Lauritzen1996, Koller2009}, this is not true when dealing with non-Gaussian data.  More specifically,  in the case of non-Gaussian data,  zero elements in the precision matrix do not necessarily imply conditional independence but just zero partial correlation \citep{Baba2004}.

By questioning the appropriateness of using the inverse of the  covariance matrix to capture the complete dependence structure in a non-Gaussian setting, we contribute to dependence literature and provide further discussion.  More precisely,  in this research, we provide different definitions of a generalized precision matrix (GPM), which hold for a large class of distributions by building on the local dependence function (LDF) defined by  \cite{Holland1987}.

% Furthermore, considering whether short selling is allowed or forbidden can also play a vital role in the estimation process \citep{pantaleo2011improved}.
%\cite{santos2019disentangling} tried to disentangle the diagonal and off-diagonal elements in the covariance estimation and found,  using a  mean-variance portfolio approach,  that the presence of non-zero off-diagonal elements in the covariance matrix often decrease the overall levels of mean return. 
%

Since its introduction, the LDF  has received attention as it better allows to capture hidden structures of dependence compared to a single scalar  \citep{JonesKoch2003}. While  \cite{Holland1987} define the LDF as the mixed partial derivative of the logarithmic density,  in 1996,  \citeauthor{Jones1996}  proposed an alternative motivation for the LDF starting from the correlation curve, which was previously developed as a generalization of the Pearson correlation coefficient (see \cite{Bjerve1993,  Doksum1994,  Blyth1994b, Blyth1994a}).   In his work, \cite{Jones1996} localized the linear correlation coefficient using the kernel method,  which then allowed to estimate the LDF. Then, \cite{Jones1998}  showed that the LDF is constant for the bivariate normal distribution.   Additionally, \cite{Bairamov2000} developed a local dependence measure as a localized version of the Galton correlation coefficient, which has been further motivated and extended to an application to 
the elliptically symmetric distributions by \cite{Nadarajah2003b}. This has then been adapted to the bivariate extreme value distributions by \cite{Nadarajah2003a}. Motivated by \cite{Jones1996, Jones1998}, \cite{Bairamov2003}  develop a LDF based on the regression concept and discuss its properties focusing on many bivariate distributions, including normal,  Farlie Gumbel Morgenstern,  bivariate exponential conditionals,  and Gumbel's bivariate exponential distribution. The LDF has also been considered in graphical models; for example, \cite{Whittaker2009} used the LDF as a measure of interaction while \cite{Capitanio2003} focused on the conditional independence graphs for skew-normal variables. Recently,  \cite{Morrison2017}  and  \cite{Spantini2018} used the absolute values of the LDF for modeling non-Gaussian graphical models and defining a GPM.  

 Here, we contribute to the literature by  adding alternative definitions to the GPM that we discuss in the context of the most appropriate application depending on the aim of the researcher.  Additionally, we apply them  to the multivariate t-Student distribution. This is especially valuable as generally asset returns do not follow a Gaussian distribution,  which is an assumption of the Markowitz model, and are often characterized by fat tails, which can increase the potential errors in estimation. In particular, we show that conditional independence between elements of $\underline{X}$  depends not just on the inverse of the covariance matrix but also on additional elements. Furthermore, we provide a graphical representation of the LDF to show that degree and direction of dependence varies in different plane regions.
Additionally, an empirical application of the minimum-variance portfolio using daily logarithmic returns of constituents of the  S\&P 100 and also the Fama and French industry portfolios are considered.   As this approach depends only on the precision matrix,  it allows us to show the effect  of using different GPMs as the input parameter. Following a rolling window approach, we show that often our portfolio computed using the newly defined GPMs has statistically significantly lower out-of-sample variance versus portfolios using the inverse sample covariance.  Furthermore, we are able to show that the new GPM is more stable than the inverse covariance matrix by using the Frobenious norm between consecutive estimates. 

The paper takes the following form: Section 2 introduces a measure of dependence for continuous densities and defines different GPMs, while Section 3 applies this measure to the t-Student distribution and provides a Taylor approximation. Section 4 discusses the dependence in a bivariate case using contour plots of the LDF, while Section 5 focuses on the minimum-variance portfolio application using the S\&P 100  and  Fama and French industry data. Lastly,  Section 6 concludes. 

\section{A measure of dependence for continuous densities}\label{general}

The notion of local dependence function (LDF) was introduced by  \cite{Holland1987} and \cite{Jones1996}. While \cite{Holland1987} tried to understand the dependence between two random variables using a limiting argument for the cross product ratio of probabilities, \cite{Jones1996} obtains the same measure starting from a local (using kernel smoothing) definition of the correlation coefficient. 
 
The LDF can be defined,  as discussed by \cite{Holland1987} and \cite{Jones1996}, given two random variables $({X_1},{X_2})$ with continuous density $f \left(x_{1}, x_{2}\right)$ with support $K= \{(x_1,x_2): f(x_1,x_2) >0 \}$ as the mixed partial derivative of the logarithm of the density; such that

\begin{equation}\label{condis}
\gamma\left(x_{1}, x_{2}\right)=\frac{\partial^2}{\partial x_{1} \partial x_{2}} \log f\left(x_{1}, x_{2}\right)  \quad  \forall (x_1,x_2)\in K. 
\end{equation}

Equation (\ref{condis}) can be defined for any positive and mixed-differentiable function.   $\gamma\left(x_{1}, x_{2}\right)$  allows to identify the dependence structure including different degrees and direction of dependence by varying $x_{1}, x_{2}$.  Independence between $x_1$ and $x_2$ holds if and only if   $\gamma\left(x_{1}, x_{2}\right)=0, \forall x_{1}, x_{2}$.  
We refer to Section \ref{Sim} for the discussion of dependence, as we visualize Equation (\ref{condis}) in Figure (\ref{fig:LDF1}).

Consider now what could be a possible extension to the multivariate case. Given a $d$-dimensional multivariate vector $\underline{X}=[X_1,X_2, \cdots, X_d]^{\top}$  with continuous and strictly positive density $f \left(x_1,\cdots,x_d \right)$ with support  $K= \{(x_1,\cdots,x_d): f(x_1,\cdots,x_d >0 \}$,  one can define $\gamma(x_p,x_q)$ conditionally on the remaining variables $\underline{X}_{\setminus p,q}$, where  $(p,q )= 1, \dots , d$; such that

\begin{equation}\label{LDF2}
    \gamma(x_p,x_q|\underline{x}_{\setminus p,q})= \frac{\partial^2}{\partial x_p \partial x_q}\log f(\underline{x})   \quad  \forall (\underline{x})\in K. 
\end{equation}

Here, we can observe conditional independence of $x_p$ and $x_q$ given all remaining variables $\underline{X}_{\setminus p,q}$ if and only if  $    \gamma(x_p,x_q|\underline{x}_{\setminus p,q})=0, \forall ({x_p},x_q)$.   This has recently been shown by \cite{Whittaker2009} in the context of graphical models.

By taking expectation with respect to the conditioning variables, one can define an average measure of dependence as 
\begin{equation}\label{ALDF}
    \overline{\gamma}(x_p,x_q) = \operatorname{E}_{\underline{X}_{\setminus p,q}} \left [\gamma(x_p,x_q|\underline{X}_{\setminus p,q}).\right ] 
\end{equation}

While Equations \eqref{LDF2} and  \eqref{ALDF} allow us to understand the direction and degree of dependence by varying $(x_p,x_q)$,  defining a GPM, say $ {\Omegabold}$, requires a further expectation step that could be defined in different ways.   

Let the first one be defined by simple expectation, i.e.
\begin{equation}\label{GPM}
\begin{aligned}
    {\Omegabold}&=& -\operatorname{E}_{X_p,X_q} \left[ \overline{\gamma}(X_p,X_q)\right]=  {-}\operatorname{E}_{\underline{X}}\left[ \frac{\partial^2}{\partial x_p \partial x_q}\log f(\underline{X})  \right].
%%=\left[-\operatorname{E}_{\underline{X}}\left(\frac{\partial}{\partial x_p \partial x_q}\log f(\underline{X})  \right)\right]\\
        \end{aligned}
\end{equation}
Note that the minus sign allows recovering exactly the precision matrix $\Sigmabold^{-1}$  in the Gaussian case as shown in Appendix (\ref{Gaus}).
The above definition of $\Omegabold =[\omega_{pq}]$ provides an estimate of the average structure of dependence over the whole space of the distribution. Therefore, note  that if  $\omega_{pq}=0$ it does not necessarily imply conditional independence between $\{x_p$, $x_q\}$.  However, the converse is true, so  if a pair of $\{x_p$, $x_q\}$ are conditional independent, then $\omega_{pq}=0$.   Even if we cannot clearly read the conditional independence from $\Omegabold$, it allows us to focus on analyzing the interactions.   

Additionally, one could consider decomposing  $\Omegabold$  over different regions defined by the  couple $\{x_p$, $x_q\}$. For example, to analyze the strength of dependence on the tails one could define a region $A_t=\{ (x_p, x_q): x_p^2+x_q^2 \geq t \}$ and then compute 
    \begin{equation}\label{GPM_TAILS}
    \Omegabold_{A_t}=- \left[ \operatorname{E}_{X_p,X_q}  \left[ \overline{\gamma}(X_p,X_q)\right] \mathbb{I}_{A_t}(X_p,X_q)\right].
    \end{equation}
If one is interested in obtaining a precision matrix which provides null elements if and only if conditional independence is present, one needs to consider taking the expectation of the absolute value, i.e.    
\begin{equation}\label{GPM_Abs}
    {\Omegabold_{Abs}}= \operatorname{E}_{X_p,X_q} \left[ |\overline{\gamma}(X_p,X_q)| \right]= \operatorname{E}_{\underline{X}}\left(\left| \frac{\partial^2}{\partial x_p \partial x_q}\log f(\underline{X}) \right| \right).
    %  {\Omegabold_{Abs}}=%\left[\Omegabold_{pq}\right]= \left[\operatorname{E}_{x_p,x_q} \left(\left| \overline{\gamma}(x_p,x_q)\right|\right)\right]=\left[\red{-\operatorname{E}_{\underline{X}}\left(\left| \frac{\partial}{\partial x_p \partial x_q}\log f(\underline{x}) \right| \right)\right].
\end{equation}  

$\Omegabold_{Abs}$ was also used by \cite{Spantini2018} and \cite{Morrison2017}  for modeling non-Gaussian graphical models. 
In their works,  \cite{Spantini2018}  focused on investigating the low-dimensional structure of transformations between random variables, which allows characterizing complex probability distributions while 
\cite{Morrison2017} introduced a new algorithm, namely SING (Sparsity Identification in Non-Gaussian distributions),  which uses transport maps to spot sparse dependence structures in continuous and non-Gaussian probability distributions.

While both definitions of the GPM, $\Omegabold$ and $\Omegabold_{Abs}$, are useful, they differ in their appropriate application. If the aim is to estimate a conditional independence graph, one should use  $ {\Omegabold_{Abs}}$; ${\Omegabold}$, on the other hand,  allows to better understand the direction and strengths of the relationships and interaction between two variables, given the others.

In the following,  an application using the t-Student distribution is discussed.

\section{t - Student distribution and its GPMs}
It is generally accepted that financial data are characterized by excessive kurtosis which is often well captured by the multivariate t-Student distribution \citep{cont2001empirical}.

As an application of the GPM,  the starting point is the density of a $d$-variate t-Student distribution with the zero mean $\underline{\mu}$, scatter matrix $\Sigmabold^{-1}$ and $\nu$  degrees of freedom:
\begin{equation}\label{raw} f(\underline{x})=
\frac{\Gamma[(\nu+d) / 2]}{\Gamma(\nu / 2) \nu^{d / 2} \pi^{d / 2}|\Sigmabold|^{1 / 2}}\left[1+\frac{1}{\nu}(\underline{x}-\underline{\mu})^{\top}{\Sigmabold}^{-1}(\underline{x}-\underline{\mu})\right]^{-\frac{\nu+d}{2}}.
\end{equation}

Looking at the probability density function in Equation (\ref{raw}), one can notice that the tail probability decays at a polynomial rate, resulting in heavy tails \citep{Ding2016}.  Setting $k=\frac{\Gamma[(\nu+d) / 2]}{\Gamma(\nu / 2) \nu^{d / 2} \pi^{d / 2}|\Sigmabold|^{1 / 2}}$ and $\delta(\underline{x})=(\underline{x}-\underline{\mu})	^{\top}\Sigmabold^{-1}(\underline{x}-\underline{\mu})$; then \eqref{raw} reduces to
\begin{equation}\label{dis}
 f(\underline{x}) =k [1+v^{-1}\delta(\underline{x})]^{-\frac{\nu+d}{2}}.
 \end{equation}

\begin{comment}
As variance, symmetry, kurtosis, and more generally higher-order cumulants are invariant to  $\underline{\mu}$; we can,  without loss of generality, assume $\underline{\mu}=0$.
\end{comment}

% \begin{equation}\label{dens}
%\mathrm{log} f_x(\underline{x}) = \mathrm{log} \hquad k - \frac{(v+d)}{2} \hquad \mathrm{log}\hquad[1+v^{-1} \delta(\underline{x})]
%\end{equation}
%and the derivative 
%
% \begin{equation}\label{multtst}
%\begin{aligned}
%\partial_{xx'}\textrm{ln}( f_x(x)) &=- \partial_{x'} -  \frac{(v+p)}{2} \hquad v^{-1}\Sigmabold^{-1}  x \frac{1}{1+v^{-1}\delta(x)} \quad \quad \\
%&= - \frac{(v+p)}{v} \bigg ( \frac{\Sigmabold^{-1} }{1+v^{-1} \delta(x)} - \frac{\Sigmabold^{-1}  xx'\Sigmabold^{-1} }{v[1+v^{-1} \delta(x)]^2} \bigg )
%\end{aligned}
%    \end{equation}
%   

%\begin{comment}
Following \cite{Holland1987} and \cite{Jones1996} we can derive the LDF of the t-Student distribution as the second partial derivative of the logarithmic density in Equation (\ref{LDF2}) such that

\begin{equation}\label{LDF2_ausgeschrieben}
    \gamma(x_p,x_q|\underline{x}_{\setminus p,q})= -\frac{\nu+d}{\nu} \left(\frac{\Sigmabold^{-1}}{1+\nu^{-1} \delta(\underline{x})}-\frac{\Sigmabold^{-1}\underline{x}\underline{x}^{\top}\Sigmabold^{-1}}{\nu (1+\nu^{-1} \delta(\underline{x}))^2}  \right).
\end{equation}
See Appendix \ref{derivatives} for full derivation.
%\end{comment}

Taking a closer look at the Equation (\ref{LDF2}), we notice that  there are two main terms that define the dependence structure. A first term that considers only the inverse covariance matrix $\Sigmabold^{-1}$ and a second term which is a combination of $\underline{x}$ and $\Sigmabold^{-1}$ in the numerator. 
More accurately, we can see that the first term is a scaled version of $\Sigmabold^{-1}$, while the second term could be defined as an index accounting for the thickness of the tails of the distribution. 
If $\underline{x} \rightarrow 0$,  positioned at the center of the distribution, considering  $\Sigmabold^{-1}$ could still give a reliable estimate as it is very similar to the Gaussian case (see Appendix \ref{Gaus} for complete derivation); however, moving away from the center, additional elements should be considered. 
   
%If we compare this to the Gaussian case with $\underline{x} \sim \mathcal{N}(0, \boldsymbol{\Sigmabold})$
%
%\begin{equation}
%p(\underline{x})=\frac{1}{\sqrt{\operatorname{det}(2 \pi \boldsymbol{\Sigmabold})}} \exp \left[-\frac{1}{2}{\underline{x}}^{\top} \boldsymbol{\Sigmabold}^{-1}\underline{x}\right]
%\end{equation}
%  
%  \begin{equation}\label{gaus}
%\begin{aligned}
%\frac{\partial}{\partial \underline{x}} &=-p(\underline{x}) \boldsymbol{\Sigmabold}^{-1}(\underline{x}) \\
%\frac{\partial^{2} }{\partial \underline{x} \partial \underline{x}^{\top}} &=p(\underline{x})\left(\boldsymbol{\Sigmabold}^{-1}\underline{x}\underline{x}^{\top}{\boldsymbol{\Sigmabold}}^{-1}-\boldsymbol{\Sigmabold}^{-1}\right)
%\end{aligned}
%\end{equation}

\subsection{A Taylor Expansion}
Approximating a function as a polynomial is often simpler than considering the exact function.   In the following, we expand log $f(\underline{x})$ around $\delta(\underline{x})=0$ under the condition $|\delta(\underline{x})|<1$. We aim to define an approximate  GPM that could be simply estimated and which could give more insights into the structure of the GPM. By expanding the logarithm of Equation (\ref{dis}) up to the third order, we get

    \begin{equation}\label{der}
M[\log f(\underline{x})]  =-\frac{\nu+d}{2}  \left[\nu^{-1}\delta(\underline{x})-  \frac{1}{2}\nu^{-2} \delta(\underline{x})^{2}  + \frac{1}{3}\nu^{-3} \delta(\underline{x})^{3} - O(x)^4 \right].
    \end{equation}

%\begin{comment}
%By taking the expectations of the derivatives in Equation (\ref{der}) gives %us an understanding of the subtle features and interconnections among the %existing indexes among different order cumulants.  All additional %information on the computation can be found in Appendix Section %\ref{Appendix}.
%Notice that we need $\delta(\underline{x})<1$ for the convergence of the %series.Then for $\delta(\underline{x})<1$,  we can define the LDF as 
% \end{comment}   
 
 By taking derivatives, we have an approximate expression for the LDF as

\begin{equation}\label{LDF2_Tay}
\begin{aligned}
    \gamma(x_p,x_q|\underline{x}_{\setminus p,q})
 \simeq &-\frac{\nu+d}{2}  \bigg [\nu^{-1}\delta^{\prime \prime}(\underline{x}) \left[1-\nu^{-1}\delta(\underline{x})+\nu^{-2}\delta^{2}(\underline{x})\right]\\
&-\nu^{-2}\delta^{\prime}(\underline{x})\delta^{\prime}(\underline{x})^{\top} \left[1+2\nu^{-1}\delta(\underline{x})\right]\bigg ].
\end{aligned}
\end{equation}

Letting $Y$ be the standardized version of the random variable $X$, i.e. $\underline{Y}=\Sigmabold^{-1/2}(\underline{X}-\underline{\mu})$. Then by taking the expectation with respect to $\underline{X}$,  we can define another GPM,  $\Omegabold_{Taylor}$, as

%\begin{equation}\label{GPM_taylor}
%\begin{aligned}
%\Omegabold_{Taylor}&= \left[\operatorname{E}_{x_p,x_q} \left( \overline{\gamma}(x_p,x_q)\right)\right]=\left[-\operatorname{E}_{\underline{X}}\left(\frac{\partial}{\partial x_p \partial x_q}M[\log f(\underline{x})]  \right)\right].\\
%\end{aligned}
%\end{equation}

\begin{equation}\label{GPM_TAYLOR_EST}
\begin{aligned}
  \widehat{\Omegabold}_{Taylor}&
&\simeq -(\nu+d) \bigg[\nu^{-1}\Sigmabold^{-1}c +4\nu^{-3} \Sigmabold^{-1 / 2}\left(\textbf{K}(Y)+(d+2) \mathbf{I_{d}}\right) \Sigmabold^{-1 / 2}\bigg ] \\
\end{aligned}
\end{equation}

where \textbf{K}(Y) is the \cite{Mori1994} matrix of kurtosis (see e.g. example 9 in \cite{Jammalamadaka2021}) and the constant $c$ is equal to   $[-1 +2\nu^{-1}+\nu^{-1}(d) -\nu^{-2}\operatorname{tr} \left(	\textbf{K}(Y)+(d+2) \mathbf{I_{d}}\right)]$.  Overall, we find that this expansion boils down to  two components: the inverse covariance matrix times a constant ($\nu^{-1}c \Sigmabold^{-1}$) and a second part which relies on the matrix of kurtosis $\textbf{K}(Y)$.
See Appendix \ref{exp} for complete derivation.

\subsection{Estimation of the GPM for the t-Student distribution}
While in Section  \ref{general}  we discussed the general formulas for the GPM, we now provide data estimation formulas for the case of the multivariate  t-Student distribution. 
 
  Therefore, given a $d$-dimensional multivariate random sample  $\underline{X_1},  \cdots, \underline{X_n}$  and given a consistent estimate of $\hat{\Sigmabold}^{-1}$ and $\hat{\nu}$, consistent estimation of ${\Omegabold}$  for the symmetric t-Student distribution can simply be obtained by computing the empirical estimate as

\begin{equation}\label{GPM_EST}
\begin{aligned}
   \widehat{\Omegabold}
    &=&\frac{\hat{\nu}+d}{\hat{\nu}} \frac{1}{n} \sum_{i=1}^n \left[ \frac{\widehat{\Sigmabold}^{-1}}{1+\hat{\nu}^{-1} \delta(\underline{X}_i)}-\frac{\widehat{\Sigmabold}^{-1}\underline{X}_i\underline{X}_i^{\top}\widehat{\Sigmabold}^{-1}}{\hat{\nu} (1+\hat{\nu}^{-1} \delta(\underline{X_i}))^2}  \right].
        \end{aligned}
\end{equation}

%$   \widehat{\Omegabold}$ seems best suited to examine the dependence strength and direction over the different regions of the space.  
Additionally, one could consider decomposing the single elements of $   \widehat{\Omegabold}$  over different regions defined by the (single) couple $x_p, x_q$. For example, to analyze the strength of dependence on the tails one could define a region $A_t=\{ (x_p, x_q): x_p^2+x_q^2 \geq t \}$ and then compute 
    \begin{equation}
    \widehat{{\Omegabold}}_{A_t}=-\left[  \frac{1}{n} \sum_{i=1}^n \left[ \frac{\partial^2}{\partial x_p \partial x_q}\log f(\underline{X}_i)  \right] \mathbb{I}_{A_t}(X_p,X_q)\right].
    \end{equation}

Clearly $\widehat{\boldsymbol{\Omegabold}}=\widehat{\boldsymbol{\Omegabold}}_{A_t}+\widehat{\boldsymbol{\Omegabold}}_{\setminus{A_t}}$

While the above allows to create a better understanding of the interaction between variables,  considering the definition of $\widehat{\Omegabold}_{Abs}$ below allows to study the independence structure in the vector $\underline{X}_i$. Therefore, if one is interested in clearly reading the independence structure, one could consider

\begin{equation}\label{GPM_ABS_EST}
\begin{aligned}
   \widehat{\Omegabold}_{Abs}
    &=&\frac{\hat{\nu}+d}{\hat{\nu}} \frac{1}{n} \sum_{i=1}^n \left[ \left|  \frac{\widehat{\Sigmabold}^{-1}}{1+\hat{\nu}^{-1} \delta(\underline{X}_i)}-\frac{\widehat{\Sigmabold}^{-1}\underline{X}_i\underline{X}_i^{\top}\widehat{\Sigmabold}^{-1}}{\hat{\nu} (1+\hat{\nu}^{-1} \delta(\underline{X}_i))^2}   \right|  \right].
        \end{aligned}
\end{equation}
Here, by using the absolute values, we can find the independence structure by looking for the nulls in the matrix. Additionally, we can use these estimators as our input for the minimum-variance portfolio. Compared to the sample inverse covariance $\Sigmabold^{-1}$, these newly defined estimators take into account the fat tails of the distribution.

\section{Visualization of the LDF}\label{Sim}
In Figure (\ref{fig:LDF1}) below, we provide a plot of the LDF of the bivariate t-Student distribution with $\nu=6$ as defined previously in Equation (\ref{LDF2_ausgeschrieben}).  We include a positive and negative linear correlation of each $\rho=|0.7|$ and $\rho=|0.5|$. By providing a visualization, we aim to investigate the dependence when the degree and direction of the dependence are divergent in different plane regions.

Regardless of the choice of $\rho$ one can identify circularly symmetric contours.   Unlike the constant dependence obtained when using the sample inverse covariance matrix,  the LDF provides varying dependence in degree and direction. In order to correctly interpret the dependence, we follow \cite{JonesKoch2003} who suggested focusing on a ‘‘regional" analysis when working with contour plots of LDFs or even adding dependence maps if necessary.

 Focusing on the diagonal elements,  we find a positive association between $X_p$ and $X_q$ when $\rho=0.5,0.7$ in the first and third quadrant and a negative association when $\rho=-0.5,-0.7$ in the second and fourth quadrant. This finding is in line with \cite{Jones1996} who displays the LDF of the Cauchy distribution, and \cite{jones2002dependent} who focuses on a bivariate t-Student distribution with marginal distributions in different degrees of freedom.
   Note,  regardless of the  $\rho$  value, we find a certain symmetry on the diagonal, which is expected as we consider the LDF of the symmetric t-Student distribution. The strongest dependence, negative and positive,  is found in the center of the plots where the values of  $x_p$ and $x_q$ are small,  almost equal to null. %The dependence is diminished when the values for $X_p$ increase while the values for $X_q$ decrease and vice versa for $\rho=-0.5,-0.7$. If $\rho=0.5,0.7$, the dependence decreases if either $X_p$ and $X_q$ are symmetrically increasing or decreasing.
In Figure (\ref{fig:LDF1}), we also identify dependence on the off-diagonal, which is strongest at the center and diminishing when the values of  $x_p$ and $x_q$ are either decreasing or increasing symmetrically for a positive value of $\rho$ or behave oppositely for  $-\rho$. Dependence on the off-diagonal could be explained by the quadratic form $\delta(x)$ in Equation (\ref{LDF2_ausgeschrieben}). Our findings, therefore, agree with \cite{Holland1987} who found analyzing the Cauchy distribution that as $\gamma(\cdot)$ ‘‘changes sign in $K$ most measures of association can be inadequate or even misleading" (p. 869).

\begin{figure}[H]
\begin{subfigure}{.5\textwidth}
  \centering
\includegraphics[width=7cm]{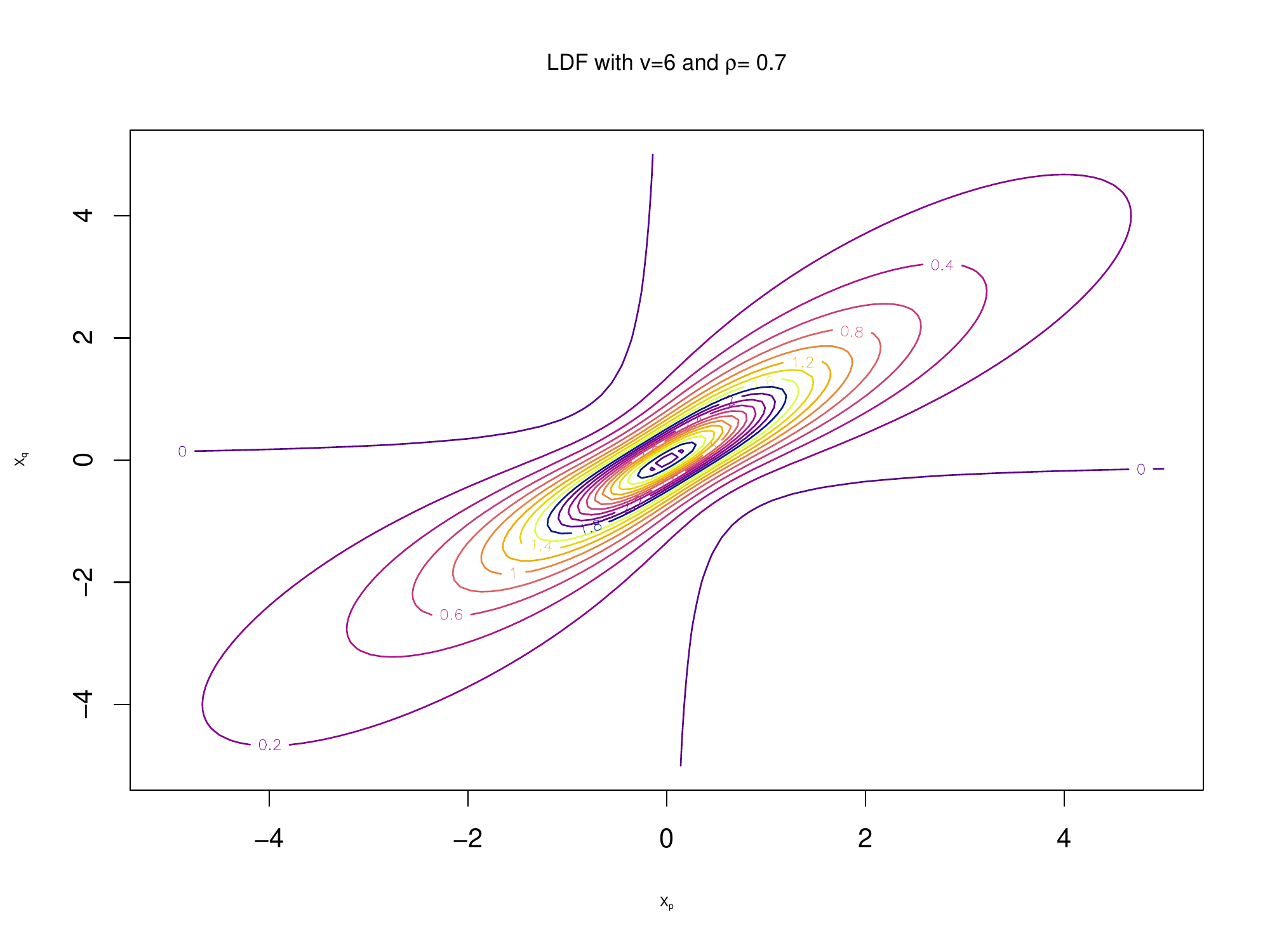}
        \caption{LDF of t-Student with $\nu=6$ and $\rho=0.7$  }    
\end{subfigure}
\begin{subfigure}{.5\textwidth}
  \centering
  % include second image
\includegraphics[width=7cm]{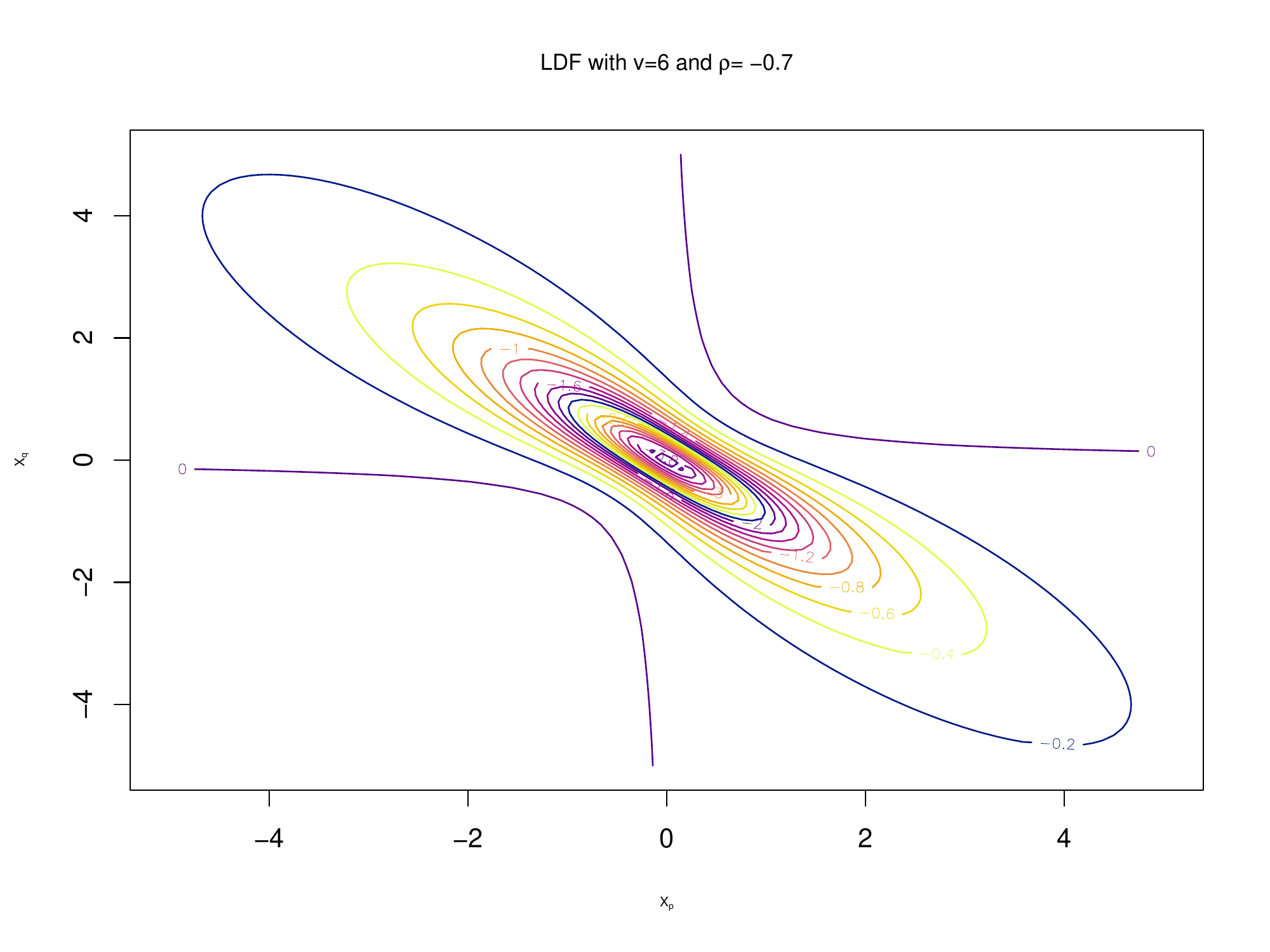}
        \caption{LDF of t-Student with $\nu=6$ and $\rho=-0.7$  }  
\end{subfigure}
\begin{subfigure}{.5\textwidth}
  \centering
\includegraphics[width=7cm]{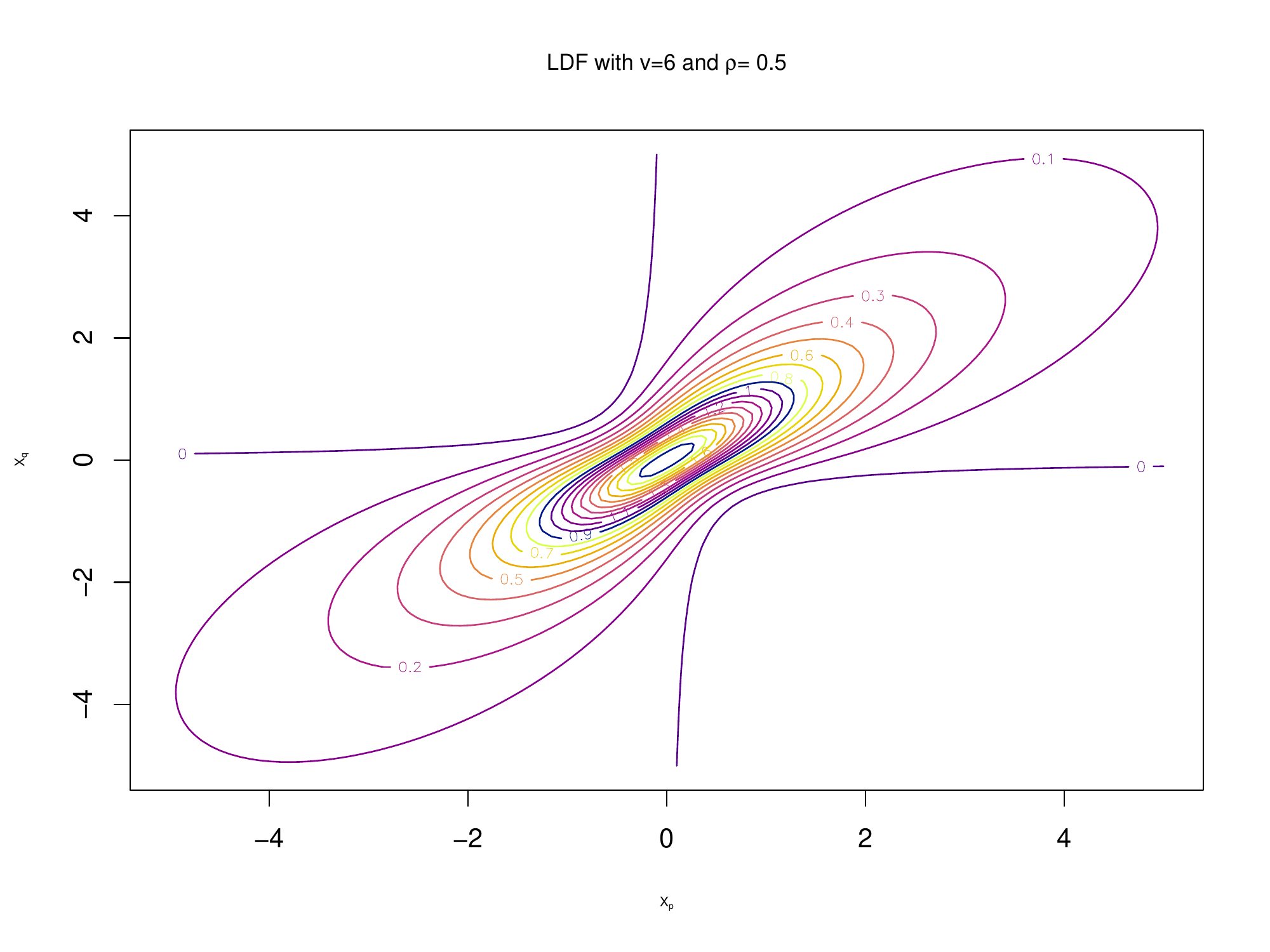}
       \caption{LDF of t-Student with $\nu=6$ and $\rho=0.5$  }    
\end{subfigure}
\begin{subfigure}{.5\textwidth}
  \centering
\includegraphics[width=7cm]{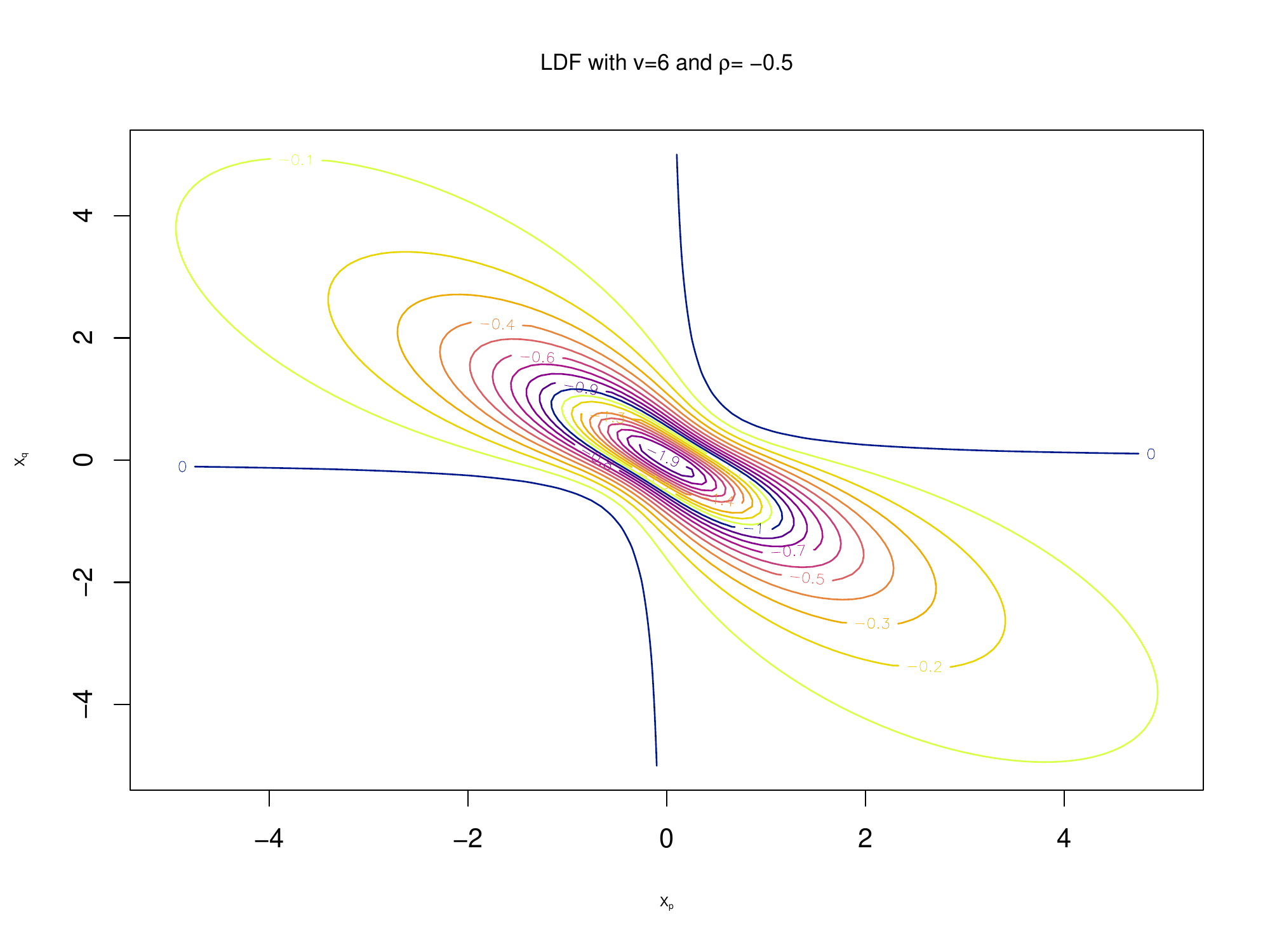}
         \caption{LDF of t-Student with $\nu=6$ and $\rho=-0.5$  } 
\end{subfigure}
\caption{Local Dependence Function of t-Student}
\label{fig:LDF1}
\end{figure}

\section{Financial Portfolio Application}
Financial portfolio optimization has attracted large attention over the years. Still today,  the Markowitz framework is the backbone of modern portfolio theory, due to its simplicity of considering only risk and return.  Here we focus on the minimum variance (MV) problem and  analyze the out-of-sample performance of the following estimators, namely $\hat{\Sigmabold}^{-1}$, $\widehat{\Omegabold}$, $\widehat{\Omegabold}_{Taylor}$, $\widehat{\Omegabold}_{Abs}$.

The minimization problem of the MV can be stated as such

\begin{equation}
\min _{\underline{w}} \hquad \underline{w}^{\top} \boldsymbol{\Sigmabold} \underline{w} \hquad   \hquad\hquad \hquad \text { s.t. } \hquad {1}^{\top} \underline{w}=1
\end{equation}
which then admits the following analytical solution
\begin{equation}
\underline{w}_{M V}=\frac{\boldsymbol{\Sigmabold}^{-1} {1}}{{1}^{\top} \mathbf{\Sigmabold}^{-1} {1}}
\end{equation}
where  $\underline{w}$ is the $n \times 1$ vector of asset weights, $\underline{1}$ a $n \times 1$ unit vector, and $\Sigmabold$ is the $n \times n$  covariance matrix. The analytical solution of the optimization $\underline{w}_{M V}$ is then the vector of weights of the optimal minimum-variance portfolio.   
As $\Sigmabold$ is unknown and moving beyond the Gaussian assumptions,  we can use the different estimators,  namely $\widehat{\Sigmabold}^{-1}$, $\widehat{\Omegabold}$, $\widehat{\Omegabold}_{Taylor}$, $\widehat{\Omegabold}_{Abs}$, as our inputs.

To test the performance of the proposed estimators,  we create a portfolio with $N$  assets and follow a rolling window approach.  We use different window sizes ($ws$), decide to re-balance the portfolios monthly ($\tau=21$), and consider various degrees of freedom $\nu$. For example, if our window size $ws=250$, we use the first 250 returns to compute the different estimates for the inverse covariance matrix using our alternative definitions of the GPM. The different estimates of the inverse covariance matrix are then used as inputs to compute the optimal weights $\underline{w}$  for the MV portfolios. The resulting portfolios are assumed to be held for the following $\tau=21$ days. Then, we roll the lock-back window forward by 21 days. By doing so, we discard the oldest 21 observations and include 21 new observations.  
 This process is repeated until the end of the time series is reached. 
Using $M$ out-of-sample returns, where  $\underline{r}_{t+1}$ is defined as the vector of returns of $\tau=21$ days,  we evaluate the resulting portfolios in terms of risk/return profile and portfolio composition.  We compute the following measures including the out-of-sample mean  ($\hat{\mu}_P$) and out-of-sample variance ($\hat{\sigma}^2_P$) (see Equations (\ref{measures1}-\ref{measures2})).
  We also report the average total turnover (TO),  the 95\% Value at Risk (VaR 95\%) computed as the empirical quantile at a 95\% confidence level, and the Sharpe Ratio assuming a risk-free  rate of zero, ($\widehat{SR}$), which serves as a risk-adjusted performance indicator (see Equations (\ref{measures3}-\ref{measures4})). We then check whether the annualized out-of-sample variances of $\widehat{\Omegabold}$, $\widehat{\Omegabold}_{Taylor}$, $\widehat{\Omegabold}_{Abs}$ are statistically significantly different to the inverse covariance matrix $\widehat{\Sigmabold}^{-1}$ using the robust performance hypothesis test by \cite{Ledoit2}.

\begin{equation}\label{measures1}
\begin{gathered}
\hat{\mu}_{P}=\frac{1}{M} \sum_{t=1}^{M} \hat{\underline{w}}_{t} \underline{r}_{t+1},
\end{gathered}
\end{equation}

\begin{equation}\label{measures2}
\begin{gathered}
\hat{\sigma}^2_{P}=\frac{1}{M-1} \sum_{i=1}^{M}\left(\hat{\underline{w}}_{t} \underline{r}_{t+1}-\hat{\mu}_{P}\right)^{2}  \\
\end{gathered}
\end{equation}

\begin{equation}\label{measures3}
\begin{gathered}
\widehat{SR}=\frac{\hat{\mu}_{p}}{\hat{\sigma}_{p}}
\end{gathered}
\end{equation}

\begin{equation}\label{measures4}
\begin{gathered}
TO =\frac{1}{M} \sum_{t=1}^{M} \sum_{i=1}^{N}\left|\hat{\underline{w}}_{i, t-1}-\hat{\underline{w}}_{i, t}\right|\\
\end{gathered}
\end{equation}

We consider two datasets to test the performance. 
More specifically, we use daily logarithmic return data of $N=80$ assets of the S\&P 100 from 03/01/2000 - 09/12/2020 and daily return data of the $N=30$ Fama and French industry portfolios of time period $t$ from 01/07/1926 - 29/10/2021.   The latter is by Kenneth French and are publicly available on his website \footnote{https://mba.tuck.dartmouth.edu/pages/faculty/ken.french/data\_library.html}.  The datasets differ on purpose in their combinations of constituents and dimensionality in order to test for robustness to the results.

In Table \ref{des} we present a summary of the descriptive statistics, where we notice that the asset return series clearly exhibit leptokurtic behavior.

\begin{table}[h]
\centering
\small
\npdecimalsign{.}
\nprounddigits{5}
\begin{tabular}{|c|c|c|n{2}{5}|n{2}{5}|n{2}{5}|n{2}{5}|c|c|c}
  \hline
 \textsc{Dataset} &  \textsc{\hspace{0.2cm}$T$}  &  \textsc{$N$}  & \textsc{\hspace{0.5cm}$\hat{\mu}$}&  \textsc{\hspace{0.3cm}$\hat{\sigma}$}  &\textsc{$\widehat{skew}$}& \textsc{$\widehat{kurt}$}& \textsc{$period$} &  \textsc{$freq.$}  \\ 
  \hline
 \textsc{FF30} &   25100 & 30 & 0.00046223213811421  & \textsc{0.014}& \textsc{0.489}  & \textsc{38.434} &  \textsc{01/07/1926 - 29/10/2021}  &  \textsc{daily} \\ 
 \textsc{SP100} & 5400  & 80 & 0.000305249505419413 & \textsc{0.022} & \textsc{-0.509}& \textsc{22.576} &  \textsc{03/01/2000 - 09/12/2020 } &  \textsc{daily}  \\ 
   \hline
\end{tabular}
\caption{The table reports a summary of the descriptive statistics for the 30 Fama and French industry portfolios (FF30) and the S\&P 100 (SP100), respectively.  Columns 1-9 report the number of observations ($T$),  the number of constituents ($k$), the averga mean ($\hat{\mu}$), the average standard deviation ($\hat{\sigma}$), the average skewness ($\widehat{skew}$) the average kurtosis ($\widehat{kurt}$) of the asset returns,  the time period ($period$) and the sampling  frequency ($freq.$).} 
\label{des}
\end{table}

Table \ref{tableSP} reports the results for $N=80$ assets of the S\&P 100  over the the complete time period.  As we re-balance the portfolio monthly ($\tau=21$), we get the following number of out-of-sample observations $M$:  with $ws=250$ we get $M=248$ and with  $ws=170$ we get $M=252$.

We find that out of all four estimators, $\widehat{\Omegabold}$ has consistently the lowest annualized out-of-sample return variance $\hat{\sigma}^2_{P}$ for all degrees of freedom and window sizes. All of these are statistically significantly different to the estimator   $\widehat{\Sigmabold}^{-1}$.
All other estimators are sometimes statistically significant; however,  their mean annualized variance is never always lower.  For window size $ws=170$ the portfolio with the largest annualized mean return,which is not the statistics to be optimized,  is always computed with the estimator  $\widehat{\Sigmabold}^{-1}$, while  portfolios built with $\widehat{\Omegabold}_{Abs}$ and  $\widehat{\Omegabold}_{Taylor}$ even encounter losses. For window size $ws=250$, it is the portfolio with estimator $\widehat{\Omegabold}_{Abs}$ which shows the highest annualized mean return while again portfolios built with $\widehat{\Omegabold}_{Taylor}$  encounter losses for $\nu=3,6$. The turnover is found to be lowest for $\widehat{\Omegabold}_{Abs}$   throughout all window sizes and degrees of freedom. The Sharpe Ratio is highest for $\widehat{\Omegabold}_{Abs}$ in window size $ws=250$ throughout all degrees of freedom while in window size $ws=170$ it is highest for $\widehat{\Sigmabold}^{-1}$. Looking at the 95\% VaR, we find that $\widehat{\Omegabold}_{Abs}$ and $\widehat{\Omegabold}_{Taylor}$ deliver the worst values  and values for   $\widehat{\Sigmabold}$ and $\widehat{\Omegabold}$ are very similar.

Overall, using the S\&P 100 data,  we show that using the GPM, more specifically,  $\widehat{\Omegabold}$,  instead of the simple inverse covariance matrix $\widehat{\Sigmabold}$  in a portfolio setting allows to significantly reduce the out-of-sample- annualized variance of such portfolio, making it valuable for investors. \\

\begin{table}[h]
\centering
\npdecimalsign{.}
\nprounddigits{5}
\begin{tabular}{|l|l|c|n{2}{5}|n{2}{5}|n{3}{5}|n{3}{5}|n{3}{6}|}\hline
\textit{ws} &Estimator&\textsc{$\nu$}  &  \textsc{\hspace{0.6cm}$\hat{\sigma}^2_{P}$}  & \textsc{\hspace{0.5cm}$\hat{\mu}_{P}$} & \textsc{\hspace{0.5cm}$\widehat{SR}$}   & \textsc{95\% VaR} & \textsc{\hspace{0.5cm}$TO$}  \\
                    \hline
 \multirow{10}{*}{170}  	&  $\widehat{\Sigmabold}^{-1}$       	& 9	& 0,037425377	& 0,196652623	& 0,064034848	&  -0,015619987 	& 2,755489538  \\   \cline{2-8} 
	&  $\widehat{\Omegabold}$        	& 9	&  \textbf{0.02969***}     	& 0,121486626	& 0,044407675	&  -0,015007085 	& 2,447267679 \\
	&  $\widehat{\Omegabold}_{Taylor}$    	& 9	& 0,037472741	& -0,023084409	& -0,00751209	&  -0,014943255 	& 0,322492786 \\
	&  $\widehat{\Omegabold}_{Abs}$  	& 9	& 0,045059192	& -0,042546646	& -0,01262622	&  -0,01700709  	& 0,103487367  \\   \cline{2-8} 
	&  $\widehat{\Omegabold}$         	& 6	&   \textbf{0.02969***}      	& 0,121406498	& 0,044381041	&  -0,015004672 	& 2,447234872 \\
	&  $\widehat{\Omegabold}_{Taylor}$      	& 6	&  1,611317081***     	& -1,70500149	& -0,084612395	&  -0,033660731 	& 5,838424185 \\
	&  $\widehat{\Omegabold}_{Abs}$ 	& 6	& 0,045059338	& -0,042547453	& -0,012626439	&  -0,017006903 	& 0,10349347  \\   \cline{2-8} 
	&  $\widehat{\Omegabold}$          	& 3	&   \textbf{0.02969***}      	& 0,121406498	& 0,044381041	&  -0,015004672 	& 2,447234872 \\
	&  $\widehat{\Omegabold}_{Taylor}$    	& 3	&  0,18041979**      	& -0,648743947	& -0,096212381	&  -0,025944256 	& 1,904050453 \\
	&  $\widehat{\Omegabold}_{Abs}$ 	& 3	& 0,045059338	& -0,042547453	& -0,012626439	&  -0,017006903 	& 0,10349347\\ \hline
 \multirow{10}{*}{250}  	&  $\widehat{\Sigmabold}^{-1}$           	& 9	& 0,026848907	& -0,007995189	& -0,003073726	&  -0,01333342  	& 1,573966863  \\   \cline{2-8} 
	&  $\widehat{\Omegabold}$            	& 9	&  \textbf{0.02554**}    	& 0,00278353	& 0,00109706	&  -0,013580945 	& 1,319721533 \\
	&  $\widehat{\Omegabold}_{Taylor}$     	& 9	&  0,035518967*     	& 0,01908888	& 0,006380431	&  -0,014737328 	& 0,130233219 \\
	&  $\widehat{\Omegabold}_{Abs}$   	& 9	&  0,040248028**     	& 0,020687398	& 0,00649581	&  -0,016349389 	& 0,064909864  \\   \cline{2-8} 
	&  $\widehat{\Omegabold}$             	& 6	&  \textbf{0.02554**}     	& 0,002866658	& 0,001129891	&  -0,013580412 	& 1,319677214 \\
	&  $\widehat{\Omegabold}_{Taylor}$     	& 6	&  0,122358241***     	& -0,209285848	& -0,037689726	&  -0,029732373 	& 1,109746 \\
	&  $\widehat{\Omegabold}_{Abs}$ 	& 6	&  0,040247635 **    	& 0,020698874	& 0,006499445	&  -0,016349117 	& 0,064912137  \\   \cline{2-8} 
	&  $\widehat{\Omegabold}$          	& 3	&  \textbf{0.02554**}     	& 0,002866658	& 0,001129891	&  -0,013580412 	& 1,319677214 \\
	&  $\widehat{\Omegabold}_{Taylor}$     	& 3	&  0,0703697***       	& -0,083315414	& -0,019784826	&  -0,024732465 	& 0,722743472 \\
	&  $\widehat{\Omegabold}_{Abs}$ 	& 3	&  0,040247635**     	& 0,020698874	& 0,006499445	&  -0,016349117 	& 0,064912137 \\
	\hline
\end{tabular}%
\caption{The table reports the out-of-sample risk and return measures of $N=80$ assets of the S\&P 100 considering different window sizes different degrees of freedom and re-balancing the portfolio every month over the period from 03/01/2000- 09/12/2020. Reported are from Column 1-8: the window size, the estimators,  the degrees of freedom, the annualized out-of-sample  variance, the annualized out-of-sample mean,  out-of-sample Sharpe Ratio, 95\% VaR and the average total turnover. Furthermore, we report the significance for the test of the difference in the annualized variance with regard to $\widehat{\Sigma}^{-1}$, at the 10\%, 5\% and 1\% level with *,**,***, respectively. The values in bold indicate the lowest annualized variance for that window size and degrees of freedom.} 
\label{tableSP}
\end{table}

\begin{table}[h]
\centering
\npdecimalsign{.}
\nprounddigits{5}
\begin{tabular}{|l|l|c|n{2}{5}|n{2}{5}|n{3}{5}|n{3}{5}|n{3}{6}|}\hline
\textit{ws} &Estimator&\textsc{$\nu$}  &  \textsc{\hspace{0.6cm}$\hat{\sigma}^2_{P}$}  & \textsc{\hspace{0.5cm}$\hat{\mu}_{P}$} & \textsc{\hspace{0.5cm}$\widehat{SR}$}   & \textsc{95\% VaR} & \textsc{\hspace{0.5cm}$TO$}  \\
                    \hline
 \multirow{10}{*}{170}  	&  $\widehat{\Sigmabold}^{-1}$   	& 9	& 0,01006902	& 0,061318753	& 0,038494568	&  -0,008465005 	& 1,001249654  \\ \cline{2-8}  	
	&  $\widehat{\Omegabold}$          	& 9	&   \textbf{0.00977}    	& 0,081453505	& 0,051916253	&  -0,008367826 	& 0,871574911 \\	
	&  $\widehat{\Omegabold}_{Taylor}$ 	& 9	&  0,013151955***     	& 0,132138924	& 0,07258307	&  -0,011236034 	& 0,411164585 \\	
	&  $\widehat{\Omegabold}_{Abs}$    	& 9	&  0,024252004***     	& 0,207233858	& 0,083827521	&  -0,013975935 	& 0,068163094  \\ \cline{2-8}  	
	&  $\widehat{\Omegabold}$          	& 6	&   \textbf{0.00977}       	& 0,0816185	& 0,052038431	&  -0,008379053 	& 0,870109071 \\	
	&  $\widehat{\Omegabold}_{Taylor}$ 	& 6	&  0,019822444***     	& 0,181406286	& 0,081165872	&  -0,013029077 	& 0,136381911 \\	
	&  $\widehat{\Omegabold}_{Abs}$    	& 6	&  0,024251587***     	& 0,20725	& 0,083834772	&  -0,013976286 	& 0,06792186  \\ \cline{2-8}  	
	&  $\widehat{\Omegabold}$          	& 3	&   \textbf{0.00977}       	& 0,0816185	& 0,052038431	&  -0,008379053 	& 0,870109071 \\	
	&  $\widehat{\Omegabold}_{Taylor}$ 	& 3	&  0,026885318***     	& 0,217136182	& 0,083420797	&  -0,014527992 	& 0,126138523 \\	
	&  $\widehat{\Omegabold}_{Abs}$    	& 3	&  0,024251587***     	& 0,20725	& 0,083834772	&  -0,013976286 	& 0,06792186  \\ \hline	
 \multirow{10}{*}{250}    	&  $\widehat{\Sigmabold}^{-1}$     	& 9	&  \textbf{0.01073}     	& -0,003178153	& -0,001932803	&  -0,008811656 	& 0,682293022  \\ \cline{2-8}  	
	&  $\widehat{\Omegabold}$          	& 9	& 0,010784557	& 0,005392054	& 0,003270791	&  -0,009099895 	& 0,598381314 \\	
	&  $\widehat{\Omegabold}_{Taylor}$ 	& 9	&  0,015436684***     	& 0,002890972	& 0,001465773	&  -0,011588287 	& 0,265538479 \\	
	&  $\widehat{\Omegabold}_{Abs}$    	& 9	&  0,026773134***     	& 0,006069573	& 0,002336729	&  -0,015716698 	& 0,046914191  \\ \cline{2-8}  	
	&  $\widehat{\Omegabold}$          	& 6	& 0,010773447	& 0,005925428	& 0,003596186	&  -0,009066602 	& 0,597344031 \\	
	&  $\widehat{\Omegabold}_{Taylor}$ 	& 6	&  0,022909224***     	& 0,006443615	& 0,002681786	&  -0,013954227 	& 0,084838752 \\	
	&  $\widehat{\Omegabold}_{Abs}$    	& 6	&  0,026771844***     	& 0,006098018	& 0,002347736	&  -0,015715106 	& 0,046723627  \\ \cline{2-8}  	
	&  $\widehat{\Omegabold}$          	& 3	& 0,010773447	& 0,005925428	& 0,003596186	&  -0,009066602 	& 0,597344031 \\	
	&  $\widehat{\Omegabold}_{Taylor}$ 	& 3	&  0,029141287***     	& 0,009093228	& 0,003355548	&  -0,016479027 	& 0,086009272 \\	
	&  $\widehat{\Omegabold}_{Abs}$    	& 3	&  0,026771844***     	& 0,006098018	& 0,002347736	&  -0,015715106 	& 0,046723627\\	
    \hline 
\end{tabular}%
\caption{The table reports the out-of-sample risk and return measures of $N=30$ Fama and French industry portfolios considering different window sizes, different degrees of freedom and re-balancing the portfolio every month over the period from  01/07/1926 - 29/10/2021.  Reported are from Column 1-8: the window size, the estimators,  the degrees of freedom, the annualized out-of-sample  variance, the annualized out-of-sample mean,  out-of-sample Sharpe Ratio, 95\% VaR and the average total turnover. Furthermore, we report the significance for the test of the difference in the annualized variance with regard to $\widehat{\Sigma}^{-1}$, at the 10\%, 5\% and 1\% level with *,**,***, respectively. The values in bold indicate the lowest annualized variance for that window size and degrees of freedom.} 
\label{tableFF}
\end{table}

We now repeat the analysis using the Fama and French industry portfolios. Table \ref{tableFF} reports the results for the Fama and French industry portfolios over the period of  01/07/1926 - 29/10/2021. As we re-balance the portfolio monthly ($\tau=21$)  and use different window sizes we get the following number of out-of-sample observations $M$:  with $ws=250$ we get $M=1183$ and with  $ws=170$ we get $M=1187$ observations.   Looking at the annualized out-of-sample return variance,  we find that portfolios with  $\widehat{\Omegabold}$  present the lowest variance for window size $ws=170$ and throughout all degrees of freedom. In the case of $ws=250$, we show that the out-of-sample return variance of $\widehat{\Omegabold}$  and $\widehat{\Sigmabold}^{-1}$ is almost equally small  resulting in no  statistically significant difference. These findings are different compared to the larger portfolio presented in Table \ref{tableSP}.

Nevertheless, we find that using $\widehat{\Omegabold}$ reduces the portfolio turnover, shows a higher Sharpe Ratio and   annualized mean return, compared to $\widehat{\Sigmabold}^{-1}$. However,  we notice that $\widehat{\Omegabold}_{Abs}$ and $\widehat{\Omegabold}_{Taylor}$ often have even lower turnover. However,  portfolios build with estimators $\widehat{\Omegabold}_{Abs}$ and $\widehat{\Omegabold}_{Taylor}$  never allow for the lowest variance.   Portfolios built with  $\widehat{\Omegabold}$  show the highest annualized mean return, which not the statistics to be optimized,  over all $ws=250$ and all degrees of freedom while for $ws=170$ this is true for $\widehat{\Omegabold}_{Abs}$.  The largest Sharpe Ratio is computed for portfolios using  $\widehat{\Omegabold}$ in case of window size of $ws=250$ and in case of window size of $ws=170$  it is $\widehat{\Omegabold}_{Abs}$.  Looking at the 95\% VaR we find that $\widehat{\Omegabold}_{Abs}$ and $\widehat{\Omegabold}_{Taylor}$ represent the worst values and again the portfolios build with estimator $\widehat{\Omegabold}$  and $\widehat{\Sigmabold}^{-1}$ are very similar.

 To summarize, using both datasets,  we find the portfolios using estimator $\widehat{\Omegabold}$  often show the smallest or at least a very similar annualized variance compared to portfolios using $\widehat{\Sigmabold}^{-1}$ while also often showing additional performance benefits including lower turnover, larger Sharpe Ratio and annualized mean return,. The effect is stronger when we look at larger portfolios and shorter time periods, as we have shown in Table \ref{tableSP}.

\begin{figure}[H]
\centering
\includegraphics[width=12cm]{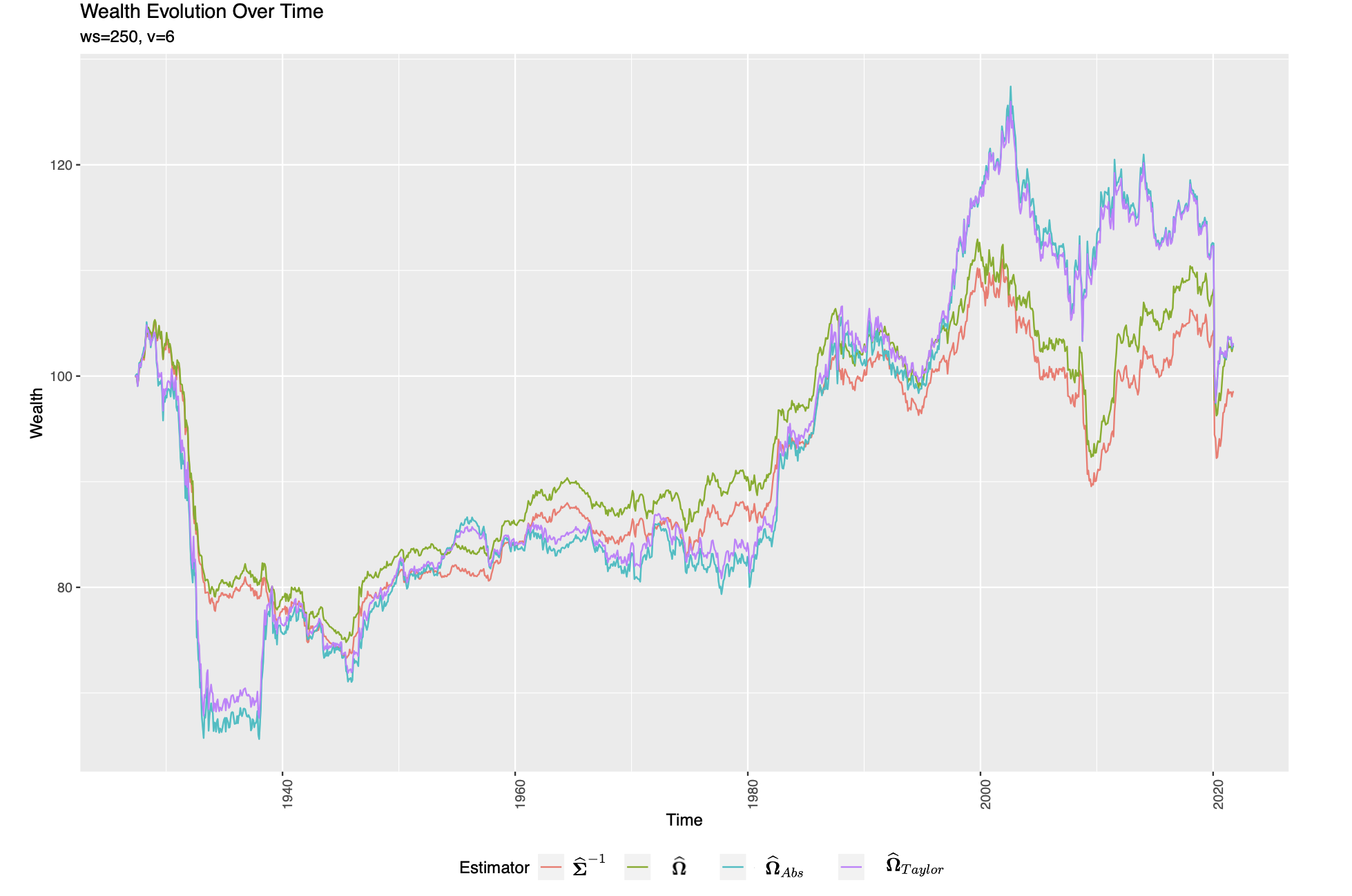}
\caption{FF30 Wealth Evolution over Rolling Window with $ws=250$ and $\nu=6$ }
\label{Rolling}
\end{figure}

Additionally, by plotting the wealth evolution of the  FF30 over the rolling window $ws=250$ with $\nu=6$ in Figure (\ref{Rolling}), we can see all portfolios develop quite similarly apart from the $\widehat{\Omegabold}_{Taylor}$ and $\widehat{\Omegabold}_{Abs}$ which are more volatile and these portfolios tend to carry higher losses and gains. This finding is as expected as the  $\widehat{\Omegabold}_{Taylor}$ is only an approximation and   $\widehat{\Omegabold}_{Abs}$ is considering only absolute values making this matrix valuable in understanding the conditional independence between random variables but not ideal in a portfolio setting. Portfolios built considering $\widehat{\Omegabold}$ often shows a slightly larger wealth than portfolios built considering $\widehat{\Sigmabold}^{-1}$. Additionally as seen from Table \ref{tableSP} and Table \ref{tableFF}, these often carry statistically significantly lower out-of-sample return variance.  Other windowsizes and degrees of freedom show show similar behavior.

In order to understand the stability of our estimators, we consider the Frobenius norm between each individual estimator $\widehat{\Omegabold}$, $\widehat{\Omegabold}_{Taylor}$, $\widehat{\Omegabold}_{Abs}$  and $\widehat{\Sigmabold}^{-1}$ for each rolling window $M$.  We compute the norm as distance between two matrices for each estimator individually.  An example for $\widehat{\Sigmabold}^{-1}$ is given in Equation (\ref{frobeq})  below.  We plot the values as boxplots for each estimator and also over the time periods to see its evolution.

\begin{equation} \label{frobeq}
\begin{aligned}
D_{F_{\hat{\Sigma}^{-1}}}&=(\widehat{\boldsymbol{\Sigma}}^{-1}_{t-1}, \widehat{\boldsymbol{\Sigma}}^{-1}_{t})=\widehat{\boldsymbol{\Sigma}}^{-1}_{t-1}, \widehat{\boldsymbol{\Sigma}}^{-1}_{t},\|_{F}=\sqrt{\operatorname{tr}\left((\widehat{\boldsymbol{\Sigma}}^{-1}_{t-1}, \widehat{\boldsymbol{\Sigma}}^{-1}_{t})(\widehat{\boldsymbol{\Sigma}}^{-1}_{t-1}, \widehat{\boldsymbol{\Sigma}}^{-1}_{t})^{\top}\right)}\\
%D_{F_{\Omega}}&=(\widehat{\boldsymbol{\Omega}}_{t-1}, \widehat{\boldsymbol{\Omega}}_{t})=\widehat{\boldsymbol{\Omega}}_{t-1}, \widehat{\boldsymbol{\Omega}}_{t},\|_{F}=\sqrt{\operatorname{tr}\left((\widehat{\boldsymbol{\Omega}}_{t-1}, \widehat{\boldsymbol{\Omega}}_{t})(\widehat{\boldsymbol{\Omega}}_{t-1}, \widehat{\boldsymbol{\Omega}}_{t})^{\top}\right)}\\
%D_{F_{\Omega}^{Abs}}&=(\widehat{\boldsymbol{\Omega}^{Abs}}_{t-1}, \widehat{\boldsymbol{\Omega}^{Abs}}_{t})=\widehat{\boldsymbol{\Omega}^{Abs}}_{t-1}, \widehat{\boldsymbol{\Omega}^{Abs}}_{t},\|_{F}=\sqrt{\operatorname{tr}\left((\widehat{\boldsymbol{\Omega}^{Abs}}_{t-1}, \widehat{\boldsymbol{\Omega}^{Abs}}_{t})(\widehat{\boldsymbol{\Omega}^{Abs}}_{t-1}, \widehat{\boldsymbol{\Omega}^{Abs}}_{t})^{\top}\right)}
\end{aligned}
\end{equation}

\begin{figure}[H]
  \centering
\includegraphics[width=12cm]{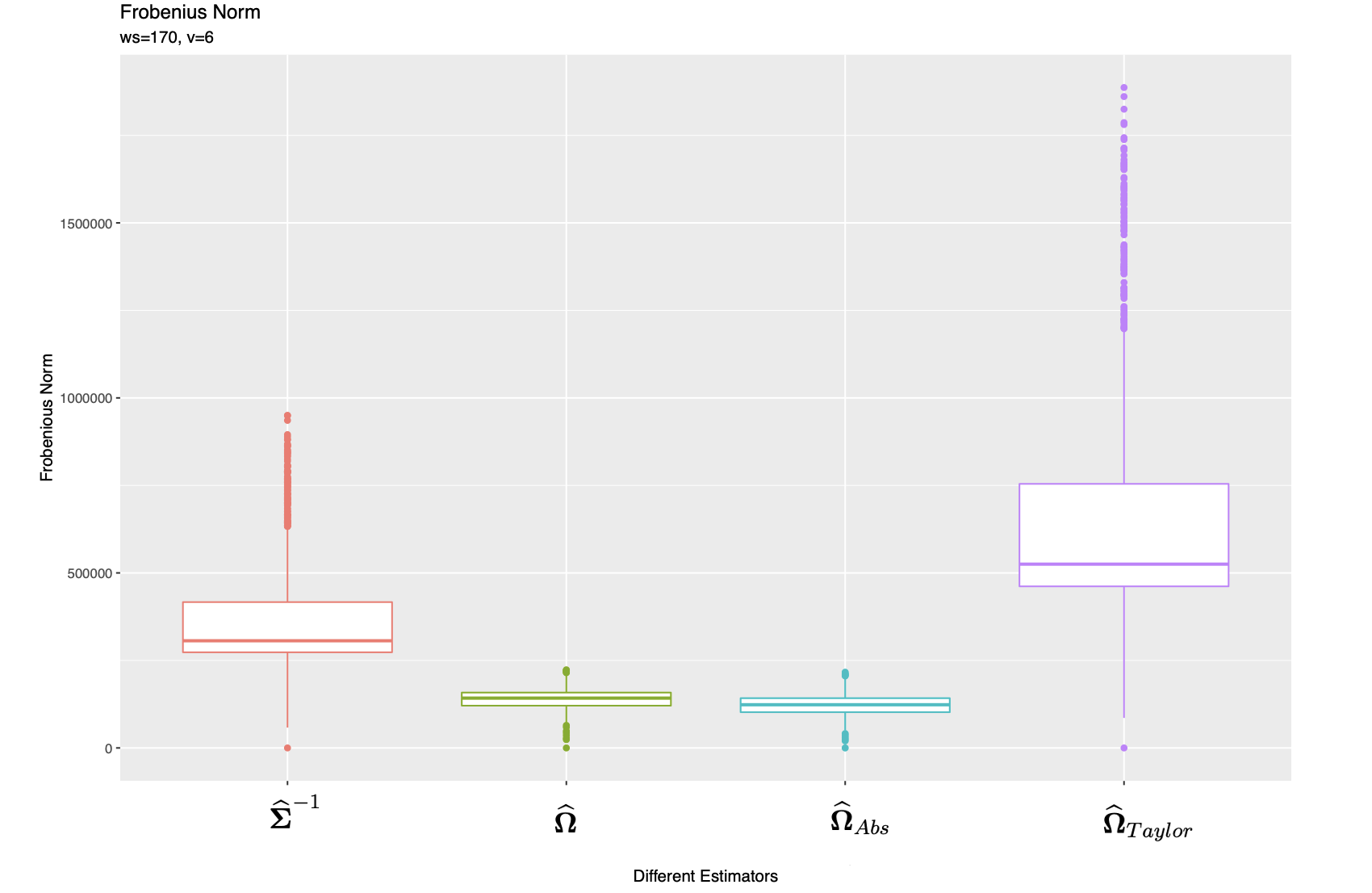}
        \caption{Frobenius norm for each estimator with $ws=170$ and $\nu=6$ of the Fama and French dataset on FF30}  
        \label{Frob1}  
\end{figure}

\begin{figure}[H]
  \centering
\includegraphics[width=12cm]{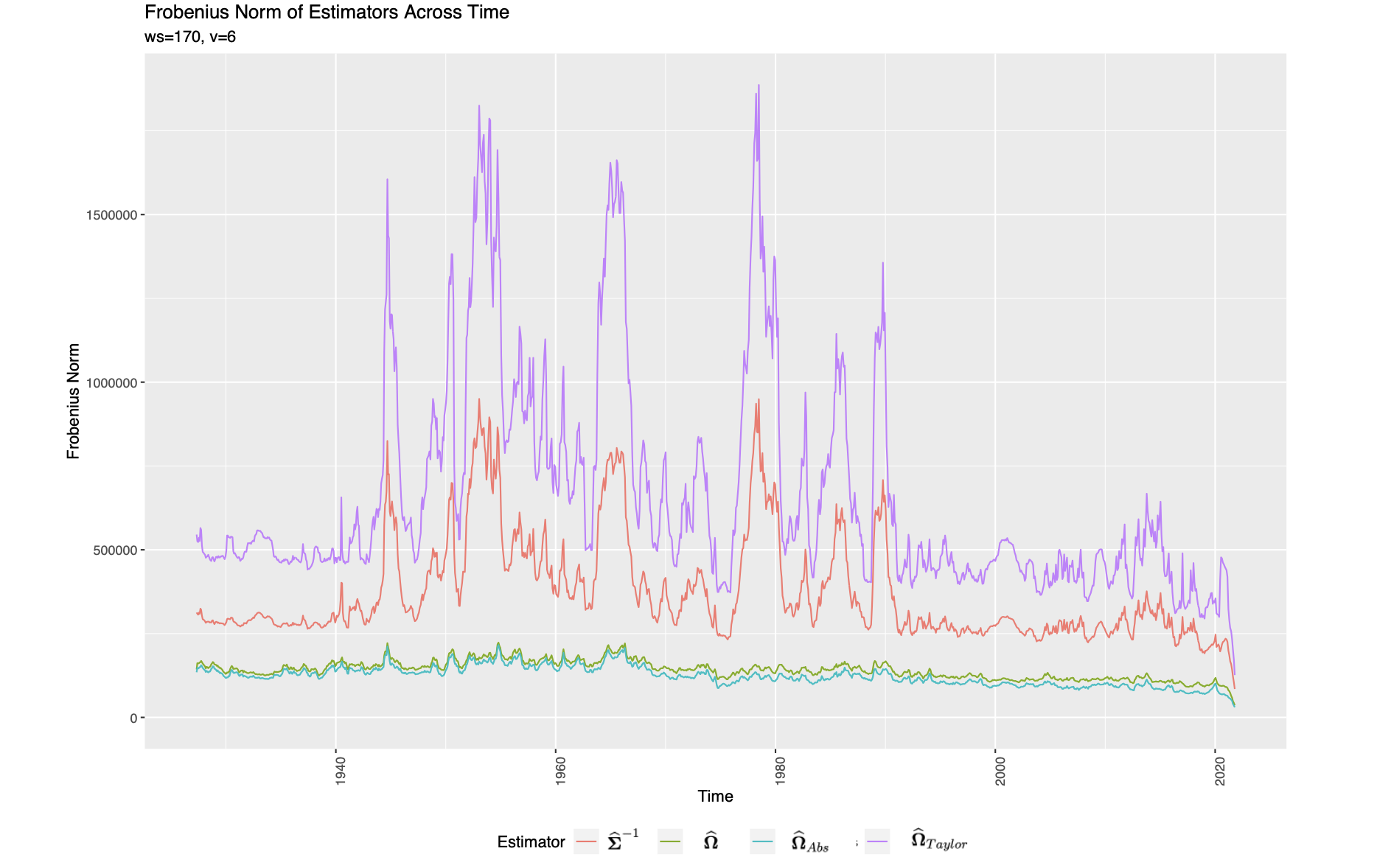}
  \caption{Frobenius norm for each estimator with $ws=170$ and $\nu=6$ for each rolling window on FF30}      
      \label{Frob2}
\end{figure}

An example using the Fama and French data with windowsize $ws=170$ and degrees of freedom $\nu=6$ is given in Figures (\ref{Frob1}-\ref{Frob2}).   Using the boxplots, we can see that $\widehat{\Omegabold}_{Taylor}$ and $\widehat{\Sigmabold}^{-1}$  seem to be estimate the most volatile estimates in consecutive periods. Additionally, throughout the whole rolling window period,  $\widehat{\Omegabold}$ seem to be more stable.  Again,  $\widehat{\Omegabold}_{Taylor}$ is very volatile and seems to be an exaggeration of $\widehat{\Sigmabold}^{-1}$. The fact that  $\widehat{\Omegabold}_{Taylor}$  is most volatile is expected as it is only an approximation.

\section{Conclusion}

In this work,  we question whether a single precision matrix, defined as the inverse covariance matrix $\Sigmabold^{-1}$, can comprehend the complete dependence structure of random assets when the Gaussian assumption fails.  

We contribute by discussing the unconditional LDF developed by \cite{Holland1987} and \cite{Jones1996} as a dependence measure for continuous densities. Then we use this information and build on \cite{Morrison2017} and \cite{Spantini2018} by providing different definitions of alternative GPMs, which hold for a large class of distributions. We discuss their most appropriate application arguing that ${\Omegabold}$  allows to better understand the strengths of the relationships and interaction between two variables, given the others, while  $ {\Omegabold_{Abs}}$ defined by  \cite{Morrison2017} and \cite{Spantini2018} is useful to estimate the conditional independence graph. Additionally, we discuss possible decomposition over different regions and introduce ${\Omegabold_{Taylor}}$ as an approximation. Considering that when working with financial data,  the assumption of Gaussianity often fails,  adapting the GPMs to different distributions is of utmost importance. As an application, we take a closer look at the t-Student distribution and provide different numerical estimations for our estimators.

By including a graphical representation of the LDF we are able to investigate the dependence when the degree and direction of the dependence are divergent in different plane regions. We agree with  \cite{Holland1987} and show that the dependence is strongest at the center and diminishes when the values of  $x_p$ and $x_q$ are either decreasing or increasing symmetrically.   This change in direction and strength might make a monotonic measure of dependence inadequate or even misleading.
 
Lastly, considering a minimum-variance portfolio application,  we are able to analyze the out-of-sample performance of our estimators. We find that for the S \&P 100,  the portfolios with $\widehat{\Omegabold}$ have a statistically significantly lower annualized variance compared to portfolios built with $\widehat{\Sigmabold}^{-1}$. When working with the Fama and French industry portfolios,  for which we consider a much longer time frame and a smaller number of assets,  we find that the outcome in terms of annualized out-of-sample variance using either $\widehat{\Omegabold}$ and $ \widehat{\Sigmabold}^{-1}$ is very similar, as their values are almost equally small which resulted in no statistically significant difference.   Nevertheless, portfolios built with  $\widehat{\Omegabold}$  show often additional performance benefits including lower turnover and larger Sharpe ratio and mean annualized return compared to portfolios built with $\widehat{\Sigmabold}^{-1}$.  Additionally,  by plotting the wealth evolution and the Frobenius norm we find that our new $\widehat{\Omegabold}$ is less volatile and more stable throughout time.

 To summarize, these findings allow us to argue that  $\widehat{\Omegabold}$ seems to be able to better capture the dependencies when dealing with t-Student data and often create portfolios with lower annualized variance and additional benefits. This is especially pronounced when dealing with a shorter time periods (here 20 years) and larger number of assets. The effect diminishes when we work with a time period of almost 100 years.  Additionally, $\widehat{\Omegabold}$ also to create more stable estimates when considering consecutive periods.
 
 Possible adaptation to the asymmetric t-Student distribution  could improve these findings and is high on the agenda.  Additionally,  not setting a priori degrees of freedom in the portfolio analysis can also be considered.

%\section{References}\label{refs}
\bibliographystyle{rQUF}
\bibliography{Paper_2_SkewedT}

\begin{thebibliography}{37}
\expandafter\ifx\csname natexlab\endcsname\relax\def\natexlab#1{#1}\fi
\providecommand{\url}[1]{\texttt{#1}}
\providecommand{\href}[2]{#2}
\providecommand{\path}[1]{#1}
\providecommand{\DOIprefix}{doi:}
\providecommand{\ArXivprefix}{arXiv:}
\providecommand{\URLprefix}{URL: }
\providecommand{\Pubmedprefix}{pmid:}
\providecommand{\doi}[1]{\href{http://dx.doi.org/#1}{\path{#1}}}
\providecommand{\Pubmed}[1]{\href{pmid:#1}{\path{#1}}}
\providecommand{\bibinfo}[2]{#2}
\ifx\xfnm\relax \def\xfnm[#1]{\unskip,\space#1}\fi
%Type = Article
\bibitem[{Baba et~al.(2004)Baba, Shibata \& Sibuya}]{Baba2004}
\bibinfo{author}{Baba, K.}, \bibinfo{author}{Shibata, R.}, \&
  \bibinfo{author}{Sibuya, M.} (\bibinfo{year}{2004}).
\newblock \bibinfo{title}{{Partial correlation and conditional correlation as
  measures of conditional independence}}.
\newblock {\it \bibinfo{journal}{Australian New Zealand Journal of
  Statistics}\/},  {\it \bibinfo{volume}{46}\/}, \bibinfo{pages}{657--664}.
%Type = Article
\bibitem[{Bairamov et~al.(2000)Bairamov, Kotz \& Kozubowski}]{Bairamov2000}
\bibinfo{author}{Bairamov, I.}, \bibinfo{author}{Kotz, S.}, \&
  \bibinfo{author}{Kozubowski, T.} (\bibinfo{year}{2000}).
\newblock \bibinfo{title}{A new measure of local dependence}.
\newblock {\it \bibinfo{journal}{Technical Report No. 362, Department of
  Statistics and Applied Probability, University of California, Santa
  Barbara}\/}, .
%Type = Article
\bibitem[{Bairamov et~al.(2003)Bairamov, Kotz \& Kozubowski}]{Bairamov2003}
\bibinfo{author}{Bairamov, I.}, \bibinfo{author}{Kotz, S.}, \&
  \bibinfo{author}{Kozubowski, T.} (\bibinfo{year}{2003}).
\newblock \bibinfo{title}{A new measure of linear local dependence}.
\newblock {\it \bibinfo{journal}{Statistics: A Journal of Theoretical and
  Applied Statistics}\/},  {\it \bibinfo{volume}{37}\/},
  \bibinfo{pages}{243--258}.
%Type = Article
\bibitem[{Bjerve \& Doksum(1993)}]{Bjerve1993}
\bibinfo{author}{Bjerve, S.}, \& \bibinfo{author}{Doksum, K.}
  (\bibinfo{year}{1993}).
\newblock \bibinfo{title}{Correlation curves: measures of association as
  functions of covariate values}.
\newblock {\it \bibinfo{journal}{The Annals of Statistics}\/},  {\it
  \bibinfo{volume}{21}\/}, \bibinfo{pages}{890--902}.
%Type = Article
\bibitem[{Black \& Litterman(1990)}]{black1990asset}
\bibinfo{author}{Black, F.}, \& \bibinfo{author}{Litterman, R.}
  (\bibinfo{year}{1990}).
\newblock \bibinfo{title}{Asset allocation: combining investor views with
  market equilibrium}.
\newblock {\it \bibinfo{journal}{Goldman Sachs Fixed Income Research}\/},  {\it
  \bibinfo{volume}{115}\/}.
%Type = Article
\bibitem[{Black \& Litterman(1992)}]{black1992global}
\bibinfo{author}{Black, F.}, \& \bibinfo{author}{Litterman, R.}
  (\bibinfo{year}{1992}).
\newblock \bibinfo{title}{Global portfolio optimization}.
\newblock {\it \bibinfo{journal}{Financial Analysts Journal}\/},  {\it
  \bibinfo{volume}{48}\/}, \bibinfo{pages}{28--43}.
%Type = Article
\bibitem[{Bloomfield et~al.(1977)Bloomfield, Leftwich \&
  Long~Jr}]{bloomfield1977portfolio}
\bibinfo{author}{Bloomfield, T.}, \bibinfo{author}{Leftwich, R.}, \&
  \bibinfo{author}{Long~Jr, J.~B.} (\bibinfo{year}{1977}).
\newblock \bibinfo{title}{Portfolio strategies and performance}.
\newblock {\it \bibinfo{journal}{Journal of Financial Economics}\/},  {\it
  \bibinfo{volume}{5}\/}, \bibinfo{pages}{201--218}.
%Type = Article
\bibitem[{Blyth(1994{\natexlab{a}})}]{Blyth1994b}
\bibinfo{author}{Blyth, S.} (\bibinfo{year}{1994}{\natexlab{a}}).
\newblock \bibinfo{title}{Karl pearson and the correlation curve}.
\newblock {\it \bibinfo{journal}{International Statistical Review/Revue
  Internationale de Statistique}\/},  {\it \bibinfo{volume}{62}\/},
  \bibinfo{pages}{393--403}.
%Type = Article
\bibitem[{Blyth(1994{\natexlab{b}})}]{Blyth1994a}
\bibinfo{author}{Blyth, S.~J.} (\bibinfo{year}{1994}{\natexlab{b}}).
\newblock \bibinfo{title}{Measuring local association: an introduction to the
  correlation curve}.
\newblock {\it \bibinfo{journal}{Sociological Methodology}\/},  {\it
  \bibinfo{volume}{24}\/}, \bibinfo{pages}{171--197}.
%Type = Article
\bibitem[{Capitanio et~al.(2003)Capitanio, Azzalini \&
  Stanghellini}]{Capitanio2003}
\bibinfo{author}{Capitanio, A.}, \bibinfo{author}{Azzalini, A.}, \&
  \bibinfo{author}{Stanghellini, E.} (\bibinfo{year}{2003}).
\newblock \bibinfo{title}{{Graphical models for skew‐normal variates}}.
\newblock {\it \bibinfo{journal}{Scandinavian Journal of Statistics}\/},  {\it
  \bibinfo{volume}{30}\/}, \bibinfo{pages}{129--144}.
%Type = Article
\bibitem[{Cont(2001)}]{cont2001empirical}
\bibinfo{author}{Cont, R.} (\bibinfo{year}{2001}).
\newblock \bibinfo{title}{Empirical properties of asset returns: stylized facts
  and statistical issues}.
\newblock {\it \bibinfo{journal}{Quantitative finance}\/},  {\it
  \bibinfo{volume}{1}\/}, \bibinfo{pages}{223}.
%Type = Article
\bibitem[{DeMiguel et~al.(2009)DeMiguel, Garlappi \&
  Uppal}]{demiguel2009optimal}
\bibinfo{author}{DeMiguel, V.}, \bibinfo{author}{Garlappi, L.}, \&
  \bibinfo{author}{Uppal, R.} (\bibinfo{year}{2009}).
\newblock \bibinfo{title}{Optimal versus naive diversification: How inefficient
  is the 1/n portfolio strategy?}
\newblock {\it \bibinfo{journal}{The review of Financial studies}\/},  {\it
  \bibinfo{volume}{22}\/}, \bibinfo{pages}{1915--1953}.
%Type = Article
\bibitem[{Ding(2016)}]{Ding2016}
\bibinfo{author}{Ding, P.} (\bibinfo{year}{2016}).
\newblock \bibinfo{title}{{On the conditional distribution of the multivariate
  t distribution}}.
\newblock {\it \bibinfo{journal}{The American Statistician}\/},  {\it
  \bibinfo{volume}{70}\/}, \bibinfo{pages}{293--295}.
%Type = Article
\bibitem[{Doksum et~al.(1994)Doksum, Blyth, Bradlow, Meng \& Zhao}]{Doksum1994}
\bibinfo{author}{Doksum, K.}, \bibinfo{author}{Blyth, S.},
  \bibinfo{author}{Bradlow, E.}, \bibinfo{author}{Meng, X.-L.}, \&
  \bibinfo{author}{Zhao, H.} (\bibinfo{year}{1994}).
\newblock \bibinfo{title}{Correlation curves as local measures of variance
  explained by regression}.
\newblock {\it \bibinfo{journal}{Journal of the American Statistical
  Association}\/},  {\it \bibinfo{volume}{89}\/}, \bibinfo{pages}{571--582}.
%Type = Article
\bibitem[{Holland \& Wang(1987)}]{Holland1987}
\bibinfo{author}{Holland, P.~W.}, \& \bibinfo{author}{Wang, Y.~J.}
  (\bibinfo{year}{1987}).
\newblock \bibinfo{title}{{Dependence function for continuous bivariate
  densities}}.
\newblock {\it \bibinfo{journal}{Communications in Statistics-Theory and
  Methods}\/},  {\it \bibinfo{volume}{16}\/}, \bibinfo{pages}{863--876}.
%Type = Article
\bibitem[{Jammalamadaka et~al.(2021)Jammalamadaka, Taufer \&
  Terdik}]{Jammalamadaka2021}
\bibinfo{author}{Jammalamadaka, S.~R.}, \bibinfo{author}{Taufer, E.}, \&
  \bibinfo{author}{Terdik, G.~H.} (\bibinfo{year}{2021}).
\newblock \bibinfo{title}{{On multivariate skewness and kurtosis}}.
\newblock {\it \bibinfo{journal}{Sankhya A}\/},  {\it
  \bibinfo{volume}{83-A}\/}, \bibinfo{pages}{1--38}.
%Type = Article
\bibitem[{Jones(2002)}]{jones2002dependent}
\bibinfo{author}{Jones, M.} (\bibinfo{year}{2002}).
\newblock \bibinfo{title}{A dependent bivariate t distribution with marginals
  on different degrees of freedom}.
\newblock {\it \bibinfo{journal}{Statistics \& probability letters}\/},  {\it
  \bibinfo{volume}{56}\/}, \bibinfo{pages}{163--170}.
%Type = Article
\bibitem[{Jones(1996)}]{Jones1996}
\bibinfo{author}{Jones, M.~C.} (\bibinfo{year}{1996}).
\newblock \bibinfo{title}{{The local dependence function}}.
\newblock {\it \bibinfo{journal}{Biometrika}\/},  {\it \bibinfo{volume}{83}\/},
  \bibinfo{pages}{899--904}.
%Type = Article
\bibitem[{Jones(1998)}]{Jones1998}
\bibinfo{author}{Jones, M.~C.} (\bibinfo{year}{1998}).
\newblock \bibinfo{title}{{Constant local dependence}}.
\newblock {\it \bibinfo{journal}{Journal of Multivariate Analysis}\/},  {\it
  \bibinfo{volume}{64}\/}, \bibinfo{pages}{148--155}.
%Type = Article
\bibitem[{Jones \& Koch(2003)}]{JonesKoch2003}
\bibinfo{author}{Jones, M.~C.}, \& \bibinfo{author}{Koch, I.}
  (\bibinfo{year}{2003}).
\newblock \bibinfo{title}{Dependence maps: Local dependence in practice}.
\newblock {\it \bibinfo{journal}{Statistics and Computing}\/},  {\it
  \bibinfo{volume}{13}\/}, \bibinfo{pages}{241--255}.
%Type = Book
\bibitem[{Koller \& Friedman(2009)}]{Koller2009}
\bibinfo{author}{Koller, D.}, \& \bibinfo{author}{Friedman, N.}
  (\bibinfo{year}{2009}).
\newblock {\it \bibinfo{title}{Probabilistic graphical models: principles and
  techniques}\/}.
\newblock \bibinfo{publisher}{MIT press}.
%Type = Article
\bibitem[{Kotz \& Nadarajah(2003)}]{Nadarajah2003b}
\bibinfo{author}{Kotz, S.}, \& \bibinfo{author}{Nadarajah, S.}
  (\bibinfo{year}{2003}).
\newblock \bibinfo{title}{Local dependence functions for the elliptically
  symmetric distributions}.
\newblock {\it \bibinfo{journal}{Sankhy{\=a}: The Indian Journal of
  Statistics}\/},  {\it \bibinfo{volume}{65}\/}, \bibinfo{pages}{207--223}.
%Type = Article
\bibitem[{Kritzman et~al.(2010)Kritzman, Page \&
  Turkington}]{kritzman2010defense}
\bibinfo{author}{Kritzman, M.}, \bibinfo{author}{Page, S.}, \&
  \bibinfo{author}{Turkington, D.} (\bibinfo{year}{2010}).
\newblock \bibinfo{title}{In defense of optimization: the fallacy of 1/n}.
\newblock {\it \bibinfo{journal}{Financial Analysts Journal}\/},  {\it
  \bibinfo{volume}{66}\/}, \bibinfo{pages}{31--39}.
%Type = Book
\bibitem[{Lauritzen(1996)}]{Lauritzen1996}
\bibinfo{author}{Lauritzen, S.~L.} (\bibinfo{year}{1996}).
\newblock {\it \bibinfo{title}{Graphical models}\/}.
\newblock \bibinfo{publisher}{Clarendon Press}.
%Type = Article
\bibitem[{Ledoit \& Wolf(2004{\natexlab{a}})}]{ledoit2004honey}
\bibinfo{author}{Ledoit, O.}, \& \bibinfo{author}{Wolf, M.}
  (\bibinfo{year}{2004}{\natexlab{a}}).
\newblock \bibinfo{title}{Honey, {I} shrunk the sample covariance matrix}.
\newblock {\it \bibinfo{journal}{The Journal of Portfolio Management}\/},  {\it
  \bibinfo{volume}{30}\/}, \bibinfo{pages}{110--119}.
%Type = Article
\bibitem[{Ledoit \& Wolf(2004{\natexlab{b}})}]{ledoit2004well}
\bibinfo{author}{Ledoit, O.}, \& \bibinfo{author}{Wolf, M.}
  (\bibinfo{year}{2004}{\natexlab{b}}).
\newblock \bibinfo{title}{A well-conditioned estimator for large-dimensional
  covariance matrices}.
\newblock {\it \bibinfo{journal}{Journal of multivariate analysis}\/},  {\it
  \bibinfo{volume}{88}\/}, \bibinfo{pages}{365--411}.
%Type = Article
\bibitem[{Ledoit \& Wolf(2011)}]{Ledoit2}
\bibinfo{author}{Ledoit, O.}, \& \bibinfo{author}{Wolf, M.}
  (\bibinfo{year}{2011}).
\newblock \bibinfo{title}{{Robust performances hypothesis testing with the
  variance}}.
\newblock {\it \bibinfo{journal}{Wilmott}\/},  {\it \bibinfo{volume}{2011}\/},
  \bibinfo{pages}{86--89}.
%Type = Article
\bibitem[{Markowitz(1952)}]{markowitz1952portfolio}
\bibinfo{author}{Markowitz, H.} (\bibinfo{year}{1952}).
\newblock \bibinfo{title}{Portfolio selection}.
\newblock {\it \bibinfo{journal}{The Journal of Finance}\/},  {\it
  \bibinfo{volume}{7}\/}, \bibinfo{pages}{77--91}.
%Type = Book
\bibitem[{Meucci(2009)}]{meucci2009risk}
\bibinfo{author}{Meucci, A.} (\bibinfo{year}{2009}).
\newblock {\it \bibinfo{title}{Risk and asset allocation}\/}.
\newblock \bibinfo{publisher}{Springer Science \& Business Media}.
%Type = Article
\bibitem[{Michaud(1989)}]{michaud1989markowitz}
\bibinfo{author}{Michaud, R.~O.} (\bibinfo{year}{1989}).
\newblock \bibinfo{title}{The {M}arkowitz optimization enigma: Is optimized
  optimal?}
\newblock {\it \bibinfo{journal}{Financial Analysts Journal}\/},  {\it
  \bibinfo{volume}{45}\/}, \bibinfo{pages}{31--42}.
%Type = Article
\bibitem[{M{\'{o}}ri et~al.(1994)M{\'{o}}ri, Rohatgi \&
  Sz{\'{e}}kely}]{Mori1994}
\bibinfo{author}{M{\'{o}}ri, T.~F.}, \bibinfo{author}{Rohatgi, V.~K.}, \&
  \bibinfo{author}{Sz{\'{e}}kely, G.~J.} (\bibinfo{year}{1994}).
\newblock \bibinfo{title}{{On multivariate skewness and kurtosis}}.
\newblock {\it \bibinfo{journal}{Theory of Probability {\&} Its
  Applications}\/},  {\it \bibinfo{volume}{38}\/}, \bibinfo{pages}{547--551}.
%Type = Inproceedings
\bibitem[{Morrison et~al.(2017)Morrison, Baptista \& Marzouk}]{Morrison2017}
\bibinfo{author}{Morrison, R.~E.}, \bibinfo{author}{Baptista, R.}, \&
  \bibinfo{author}{Marzouk, Y.} (\bibinfo{year}{2017}).
\newblock \bibinfo{title}{{Beyond normality: Learning sparse probabilistic
  graphical models in the non-Gaussian setting}}.
\newblock In {\it \bibinfo{booktitle}{31st Conference on Neural Information
  Processing Systems (NIPS 2017), Long Beach, CA, USA.}\/}.
\newblock \bibinfo{note}{ArXiv preprint arXiv:1711.00950}.
%Type = Article
\bibitem[{Nadarajah et~al.(2003)Nadarajah, Mitov \& Kotz}]{Nadarajah2003a}
\bibinfo{author}{Nadarajah, S.}, \bibinfo{author}{Mitov, K.}, \&
  \bibinfo{author}{Kotz, S.} (\bibinfo{year}{2003}).
\newblock \bibinfo{title}{Local dependence functions for extreme value
  distributions}.
\newblock {\it \bibinfo{journal}{Journal of Applied Statistics}\/},  {\it
  \bibinfo{volume}{30}\/}, \bibinfo{pages}{1081--1100}.
%Type = Article
\bibitem[{Spantini et~al.(2018)Spantini, Bigoni \& Marzouk}]{Spantini2018}
\bibinfo{author}{Spantini, A.}, \bibinfo{author}{Bigoni, D.}, \&
  \bibinfo{author}{Marzouk, Y.} (\bibinfo{year}{2018}).
\newblock \bibinfo{title}{{Inference via low-dimensional couplings}}.
\newblock {\it \bibinfo{journal}{The Journal of Machine Learning Research}\/},
  {\it \bibinfo{volume}{19}\/}, \bibinfo{pages}{2639--2709}.
%Type = Article
\bibitem[{Stevens(1998)}]{stevens1998inverse}
\bibinfo{author}{Stevens, G.~V.} (\bibinfo{year}{1998}).
\newblock \bibinfo{title}{On the inverse of the covariance matrix in portfolio
  analysis}.
\newblock {\it \bibinfo{journal}{The Journal of Finance}\/},  {\it
  \bibinfo{volume}{53}\/}, \bibinfo{pages}{1821--1827}.
%Type = Book
\bibitem[{Whittaker(2009)}]{Whittaker2009}
\bibinfo{author}{Whittaker, J.} (\bibinfo{year}{2009}).
\newblock {\it \bibinfo{title}{{Graphical models in applied multivariate
  statistics}}\/}.
\newblock \bibinfo{publisher}{Wiley Publishing}.
%Type = Article
\bibitem[{Won et~al.(2013)Won, Lim, Kim \& Rajaratnam}]{won2013condition}
\bibinfo{author}{Won, J.-H.}, \bibinfo{author}{Lim, J.}, \bibinfo{author}{Kim,
  S.-J.}, \& \bibinfo{author}{Rajaratnam, B.} (\bibinfo{year}{2013}).
\newblock \bibinfo{title}{Condition-number-regularized covariance estimation}.
\newblock {\it \bibinfo{journal}{Journal of the Royal Statistical Society:
  Series B (Statistical Methodology)}\/},  {\it \bibinfo{volume}{75}\/},
  \bibinfo{pages}{427--450}.

\end{thebibliography}

\section{Appendices}

\section{Derivatives and Expectation of Gaussian distribution}\label{Gaus}

The density of a $d$-variate Gaussian distribution with the zero mean $\underline{\mu}$ and scatter matrix $\Sigmabold^{-1}$ is defined as such

\begin{equation}
\begin{aligned}
f(\underline{x})&=\frac{1}{\sqrt{(2 \pi )^d\operatorname{det}(\mathbf{\Sigma})}} \exp \left[-\frac{1}{2}(\underline{x}-\underline{\mu})^{T} \mathbf{\Sigma}^{-1}(\underline{x}-\underline{\mu})\right]\\
&= \frac{1}{\sqrt{(2 \pi )^d\operatorname{det}(\mathbf{\Sigma})}} \exp \left[-\frac{1}{2} \delta(\underline{x})\right] \\
&= k \exp \left[-\frac{1}{2} \delta(\underline{x})\right]
\end{aligned}
\end{equation}

By taking the derivatives of $   \log f(\underline{\underline{x}})$  we get 
\begin{equation}
\begin{aligned}
\frac{\partial}{\partial x } \log f(\underline{x}) =- \delta^{\prime}(\underline{x})=-2 \Sigmabold^{-1} \underline{x} \\
\frac{\partial^2}{\partial x \partial x^{\top}}  \log f(\underline{x}) =- \Sigmabold^{-1} 
    \end{aligned}
\end{equation}

Then, by defining the $GPM = -\operatorname{E}_{\underline{X}}\left(  \frac{\partial^2}{\partial x \partial x^{\top}} \log f(\underline{X}) \right)$ we get 

\begin{equation}
    \Omegabold_{Gaussian} =\Sigmabold^{-1} 
\end{equation}

\section{Derivatives of t-Student distribution } \label{derivatives}
\begin{equation}\label{dis_all}
\begin{aligned}
 f(\underline{x}) 
 &=&k [1+\nu^{-1}(\underline{x}-\underline{\mu})^{\top}{\Sigmabold}^{-1}(\underline{x}-\underline{\mu}]^{-(\nu+d)/2}\\
 &=&k [1+\nu^{-1}\delta(\underline{x})]^{-(\nu+d)/2}
\end{aligned}
 \end{equation}

 \begin{equation}\label{dens}
\log  f(\underline{x}) = \mathrm{log} \hquad k - \frac{(\nu+d)}{2} \hquad \mathrm{log}\hquad[1+\nu^{-1} \delta(\underline{x})]
\end{equation}

 \begin{equation}
\begin{aligned}
  \frac{\partial}{ \partial_{x}} \log f(\underline{x})
    &=&- \frac{(\nu+d)}{2} \nu^{-1} \Sigmabold^{-1}2\underline{x} \frac{1}{1+\nu^{-1} \delta(\underline{x})} \\
\end{aligned}
 \end{equation}

Now taking the second partial derivative we get:
\begin{equation}\label{multtst}
\begin{aligned}
\frac{\partial^2}{\partial x \partial x^{\top}} \log f(\underline{x}) 
&=& -\frac{\nu+d}{2} \left(\frac{2\nu^{-1} \Sigmabold^{-1} (1+\nu^{-1} \delta(\underline{x}))- 2\nu^{-2} \Sigmabold^{-1}\underline{x}\underline{x}^{\top}\Sigmabold^{-1} }{(1+\nu^{-1} \delta(\underline{x}))^2}\right)\\
&=& -\frac{\nu+d}{2} \left(\frac{2\nu^{-1} \Sigmabold^{-1} (1+\nu^{-1} \delta(\underline{x}))}{(1+\nu^{-1} \delta(\underline{x}))^2}- \frac{2\nu^{-2} \Sigmabold^{-1}\underline{x}\underline{x}^{\top}\Sigmabold^{-1} }{(1+\nu^{-1} \delta(\underline{x}))^2}\right)\\
&=& -\frac{\nu+d}{2} \left(\frac{2\Sigmabold^{-1} }{\nu(1+\nu^{-1} \delta(\underline{x}))}- \frac{ 2\Sigmabold^{-1}\underline{x}\underline{x}^{\top}\Sigmabold^{-1} }{\nu^2(1+\nu^{-1} \delta(\underline{x}))^2}\right)\\
&=&-\frac{\nu+d}{\nu} \left(\frac{\Sigmabold^{-1}}{1+\nu^{-1} \delta(\underline{x})}-\frac{\Sigmabold^{-1}\underline{x}\underline{x}^{\top}\Sigmabold^{-1}}{\nu (1+\nu^{-1} \delta(\underline{x}))^2}  \right).
\end{aligned}
\end{equation}

As we define $GPM = -\operatorname{E}_{\underline{X}}\left(   \frac{\partial^2}{\partial x \partial x^{\top}} \log f(\underline{X}) \right)$, then given the estimates $\hat{\nu}$ and $\widehat{\Sigmabold}$ and given a multivariate random sample $ \underline{X_1}, \cdots, \underline{X_n}$, we have

\begin{equation}
\begin{aligned}
    \widehat{\Omegabold} &=& %\red{-}-\frac{\hat{\nu}+d}{\hat{\nu}} \frac{1}{n} \sum_{i=1}^n \frac{\widehat{\Sigmabold}^{-1}}{1+\hat{\nu}^{-1} \delta(\underline{X}_i)}-\frac{\widehat{\Sigmabold}^{-1}\underline{X}_i\underline{X}_i^{\top}\widehat{\Sigmabold}^{-1}}{\hat{\nu} (1+\hat{\nu}^{-1} \delta(\underline{X}_i))^2}  \\
    \frac{\hat{\nu}+d}{\hat{\nu}} \frac{1}{n} \sum_{i=1}^n \left| \frac{\widehat{\Sigmabold}^{-1}}{1+\hat{\nu}^{-1} \delta(\underline{X})}-\frac{\widehat{\Sigmabold}^{-1}\underline{X}\underline{X}^{\top}\widehat{\Sigmabold}^{-1}}{\hat{\nu} (1+\hat{\nu}^{-1} \delta(\underline{X}))^2}  \right|.
    \end{aligned}
\end{equation}

\section{Expectations and Taylor Approximation}\label{exp}
In the following the expectation of Equation(\ref{der}) is taken. Let $Y$ be the standardized version of the random variable $X$, i.e. $\underline{Y}=\Sigmabold^{-1/2}(\underline{X}-\underline{\mu})$.

%\subsection{$\mathrm{E} \delta(X)$ }

For this note that
%    \begin{equation}
%\begin{aligned}
%\mathrm{E} \delta(X) 
%&=\mathrm{E}\left(\operatorname{tr}  X^{\prime} \Sigmabold^{-1} X\right) \\
%&=\mathrm{E}\left(\operatorname{tr}  \Sigmabold^{-1} X X^{\prime}\right) \\
%&= \operatorname{tr}  \Sigmabold^{-1} \mathrm{E}\left(X X^{\prime}\right) \\
%&=\operatorname{tr} (\Sigmabold^{-1} \Sigmabold )
%=\operatorname{tr} I_{d} 
%=d
%\end{aligned}
%    \end{equation}   

  \begin{equation}
\begin{aligned}
\mathrm{E} \delta(\underline{X}) 
&=\mathrm{E}\left(\operatorname{tr}  \underline{X}^{\top} \Sigmabold^{-1} \underline{X}\right) \\
&=\mathrm{E}\left(\operatorname{tr}  \Sigmabold^{-1} \underline{X} \underline{X}^{\top}\right) \\
&= \operatorname{tr}  \Sigmabold^{-1} \mathrm{E}\left(\underline{X} \underline{X}^{\top}\right) \\
&=\operatorname{tr} (\Sigmabold^{-1} \Sigmabold )
=\operatorname{tr} \mathbf{I_{d}}
=d
\end{aligned}
    \end{equation}   
     
where we have used the facts that, the trace (defined to be the sum of elements on the main diagonal) of a scalar is the scalar itself, $\operatorname{tr}( AB)=\operatorname{tr}(BA)$ (if products exists), the trace is a linear operator, and $\mathrm{E}( \underline{X} \underline{X}^{\top})=\Sigmabold$. We know that  $\mathrm{E} \delta(\underline{X})=\mathrm{E}\left(\underline{X}^{\top} \Sigmabold^{-1} \underline{X}\right) = $ can be re-written as $E( \underline{Y}^{\top} \underline{Y})$.

% \begin{equation}
%\begin{aligned}
%\mathrm{E} \delta(X) &=\mathrm{E}\left(\operatorname{tr} X^{\prime} \Sigmabold^{-1} X\right) 
%%&= \Sigmabold^{-1} \mathrm{E}\left(\operatorname{tr} X^{\prime} X%\right) \\
%%&=   \operatorname{tr} \Sigmabold^{-1} E(Y \Sigmabold^{1/2} Y^{\prime}  %Sigmabold^{1/2} )\\
%=  \operatorname{tr}E(Y Y^{\prime} )
%\end{aligned}
%    \end{equation}   
%
%Therefore, 
%    \begin{equation}
%\begin{aligned}
%\mathrm{E} \delta^{2}(X) &=\mathrm{E}\left(\operatorname{tr} Y^{\prime} Y Y^{\prime} Y\right) \\
%&=\mathrm{E}\left(\operatorname{tr} Y Y^{\prime} Y Y^{\prime}\right) \\
%&= \mathrm{E}\left(\operatorname{tr}  Y Y^{\prime} Y Y^{\prime}\right)\\
%&= \operatorname{tr} \left(K(Y)+(d+2) I_{d}\right)\\
%\end{aligned}   
% \end{equation}   

   Therefore, 
    \begin{equation}
\begin{aligned}
\mathrm{E} \delta^{2}(\underline{X}) &=\mathrm{E}\left(\operatorname{tr} \underline{Y}^{\top} \underline{Y} \underline{Y}^{\top} \underline{Y}\right) \\
&=\mathrm{E}\left(\operatorname{tr} \underline{Y} \underline{Y}^{\top} \underline{Y} \underline{Y}^{\top}\right) \\
&= \operatorname{tr}\left(\mathrm{E} \left(  \underline{Y} \underline{Y}^{\top} \underline{Y} \underline{Y}^{\top}\right)\right)\\
&= \operatorname{tr} \left(\textbf{K}(Y)+(d+2) \mathbf{I_{d}}\right)\\
\end{aligned}   
 \end{equation}   
   
where $\textbf{K}(Y)$ is \cite{Mori1994} matrix of Kurtosis as shown in  Example 9 in \cite{Jammalamadaka2021}. %as $K(\mathrm{Y})=\mathrm{E}\left(\mathrm{Y} \mathrm{Y}^{\prime} \mathrm{Y} \mathrm{Y}^{\prime}\right)-(d+2) \mathrm{I}_{d}=\mathrm{E}\left(\mathrm{Y}^{\prime} \mathrm{Y}\right) \mathrm{Y} \mathrm{Y}^{\prime}-(d+2) \mathrm{I}_{d}$. %We also consider a standardized vector as in  \cite{Jammalamadaka2021} stating  $\mathbf{Y}=\mathbf{\Sigmabold}^{-1 / 2}(\mathbf{X}-\boldsymbol{\underline{\mu}})$ which results in  $\Sigmabold^{-1}$ as$\frac{\frac{1}{\Sigmabold}}{\Sigmabold^{-1 / 2} \Sigmabold^{-1 / 2}}=1$T
%Taking the third expectation in Equation (\ref{3}) results in the generation $\kappa_{3}$,  which is defined as Mardia's index of skewness according to Example 1 from  \cite{Jammalamadaka2021}.   In the case of symmetric multivariate distribution it is 0.
%
%   
%   \red{If we can add tr to order the Ys above, we should add it here.}
%   
%    \begin{equation}\label{3}
%\begin{aligned}
%\mathrm{E} \delta^{3}(X)
% &=\mathrm{E}\left(\operatorname{tr} Y^{\prime} Y Y^{\prime} Y Y^{\prime} Y\right) \\
% &=\mathrm{E}\left( \operatorname{tr}  Y^{\prime} Y Y^{\prime }  Y Y^{\prime} Y\right)  
%  %&=\mathrm{E}\left(Y^{\prime} Y\right)^{3}\\
%= \kappa_{3}
%\end{aligned}
%    \end{equation}   
%  
Then continuing we get

    \begin{equation}
\begin{aligned}
\mathrm{E} \delta^{\prime}(\underline{X}) \delta^{\prime}(\underline{X})^{\top}
&=
 \mathrm{E}\left(2\Sigmabold^{-1} \underline{X} \underline{X}^{\top} 2\Sigmabold^{-1} \right)
&=2\Sigmabold^{-1} \mathrm{E}\left(\underline{X} \underline{X}^{\top}\right) 2\Sigmabold^{-1}
=4\Sigmabold^{-1}\Sigmabold \Sigmabold^{-1}
=4\Sigmabold^{-1} \\
\end{aligned}
    \end{equation}  
    
      \begin{equation}
\begin{aligned}  
\mathrm{E} \delta^{\prime}(\underline{X}) \delta^{\prime}(\underline{X})^{\top} \delta(\underline{X})
&=\mathrm{E}\left(\Sigmabold^{-1} \underline{X} \underline{X}^{\top} 2\Sigmabold^{-1} \underline{X}^{\top} 2\Sigmabold^{-1}\underline{X}\right) \\
&=4\mathrm{E}\left(\Sigmabold^{-1} \underline{X} \underline{X}^{\top} \Sigmabold^{-1} \underline{X} \underline{X}^{\top} \Sigmabold^{-1}\right) \\
&= 4\mathrm{E}\left(\Sigmabold^{-1 / 2} \underline{Y} \underline{Y}^{\top} \underline{Y} \underline{Y}^{\top} \Sigmabold^{-1 / 2}\right) \\
&  =4 \bigg (\Sigmabold^{-1 / 2}\left(\textbf{K}(Y)+(d+2) \mathbf{I_{d}}\right) \Sigmabold^{-1 / 2} 4 \bigg )
\end{aligned}
    \end{equation}

%     
%      \begin{equation}
%\begin{aligned}  
%\mathrm{E} \delta^{\prime}(x) \delta^{\prime}(x)^{\top} \delta^{2}(x)
%&=\mathrm{E}\left(\Sigmabold^{-1} X X^{\prime} \Sigmabold^{-1} X^{\prime} \Sigmabold^{-1} X X^{\prime} \Sigmabold^{-1} X\right) \\
%&=\mathrm{E}\left(\Sigmabold^{-1} X X^{\prime} \Sigmabold^{-1} X X^{\prime} \Sigmabold^{-1} X X^{\prime} \Sigmabold^{-1}\right) \\
%&= \mathrm{E}\left(\Sigmabold^{-1 / 2} Y Y^{\prime} Y Y^{\prime}  Y Y^{\prime} \Sigmabold^{-1 / 2}\right) \\
%&=\Sigmabold^{-1 / 2}\left( \mathrm{E}\left(YY^{\prime} \right)^{3}   \right) \Sigmabold^{-1 / 2}
%\\
%&=???
%\end{aligned}
%    \end{equation}   

Then,  we can use the results of this section and compute the expectation of the approximated LDF defined previously in Equation (\ref{LDF2_Tay}); such that
%
% \begin{equation}
%\begin{aligned}
%-\mathrm{E}_{\underline{X}}\left[ \frac{\partial^2}{\partial x \partial x^{\top}}  \mathrm{ln}\hquad(1+\delta(\underline{x})) \right] 
%&\simeq-\Sigmabold^{-1}+ \Sigmabold^{-1}(d)-\Sigmabold^{-1}\operatorname{tr} \left(	\textbf{K(Y)}+(d+2) \mathbf{I_{d}}\right) \\
%&+ \Sigmabold^{-1} +2 \Sigmabold^{-1 / 2}\left(	\textbf{K(Y)}+(d+2) \mathbf{I_{d}}\right) \Sigmabold^{-1 / 2} \\
%&\simeq \Sigmabold^{-1}(d)-\Sigmabold^{-1}\operatorname{tr}\left(	\textbf{K(Y)}+(d+2) \mathbf{I_{d}}\right) \\
%&+2 \Sigmabold^{-1 / 2}\left(	\textbf{K(Y)}+(d+2) \mathbf{I_{d}}\right) \Sigmabold^{-1 / 2} \\
%&\simeq \Sigmabold^{-1}(d)-\Sigmabold^{-1}\operatorname{tr}(	\textbf{K(Y)})-\Sigmabold^{-1}d(d+2)\\
%&+2 \Sigmabold^{-1 / 2}\left(	\textbf{K(Y)}+(d+2) \mathbf{I_{d}}\right) \Sigmabold^{-1 / 2} \\
%\end{aligned}
%\end{equation}

\begin{equation}
\footnotesize
\begin{aligned}
\widehat{\Omegabold}_{Taylor}
%-\mathrm{E}_{\underline{X}}\left[ \frac{\partial^2}{\partial x \partial x^{\top}}   %-\frac{\nu+d}{2} \hquad \log \hquad(1+\nu^{-1}\delta(\underline{X})) \right] 
&\simeq -\frac{\nu+d}{2}\bigg[-2\nu^{-1}\Sigmabold^{-1}+ 2\nu^{-2}\Sigmabold^{-1}(d)-2\nu^{-3}\Sigmabold^{-1}\operatorname{tr} \left(	\textbf{K}(Y)+(d+2) \mathbf{I_{d}}\right) \\
&+ 4\nu^{-2}\Sigmabold^{-1} +2\nu^{-3}4 \Sigmabold^{-1 / 2}\left(\textbf{K}(Y)+(d+2) \mathbf{I_{d}}\right) \Sigmabold^{-1 / 2}\bigg ] \\
&\simeq -(\nu+d) \bigg[-\nu^{-1}\Sigmabold^{-1}+ \nu^{-2}\Sigmabold^{-1}(d)-\nu^{-3}\Sigmabold^{-1}\operatorname{tr} \left(	\textbf{K}(Y)+(d+2) \mathbf{I_{d}}\right) \\
&+ 2\nu^{-2}\Sigmabold^{-1} +4\nu^{-3} \Sigmabold^{-1 / 2}\left(\textbf{K}(Y)+(d+2) \mathbf{I_{d}}\right) \Sigmabold^{-1 / 2}\bigg ] \\
\\
\end{aligned}
\end{equation}

\end{document}